\documentclass[12pt]{article}
\pdfoutput=1
\usepackage{ulem}
\usepackage{subfigure}
\usepackage{amssymb,amsmath}
\usepackage{graphicx}
\usepackage{color}
\usepackage[colorlinks=true
,urlcolor=blue
,citecolor=blue
,linkcolor=blue
,pagecolor=blue
,linktocpage=true
,pdfproducer=medialab
]{hyperref}
\usepackage[a4paper,width=15.2cm]{geometry}

\makeatletter \renewcommand{\@dotsep}{10000} \makeatother

\newcommand{\beq}{\begin{equation}}
\newcommand{\eeq}{\end{equation}}
\newcommand{\bea}{\begin{eqnarray}}
\newcommand{\eea}{\end{eqnarray}}

\begin{document}

\begin{center}

 {\Large\bf    Nonuniversal Gaugino Masses and Muon $g-2$
  } \vspace{1cm}

{\large     Ilia Gogoladze\footnote{ E-mail: ilia@bartol.udel.edu\\
\hspace*{0.6cm} On  leave of absence from: Andronikashvili Institute
of Physics,  Tbilisi, Georgia.}, Fariha Nasir\footnote { E-mail: fariha@udel.edu },   Qaisar Shafi\footnote{ E-mail:
shafi@bartol.udel.edu} and Cem Salih $\ddot{\rm U}$n\footnote{
E-mail: cemsalihun@bartol.udel.edu}} \vspace{.5cm}

{\baselineskip 20pt
Bartol Research Institute, Department of Physics and Astronomy, \\
University of Delaware, Newark, DE 19716, USA  } \vspace{.5cm}

\vspace{1.5cm}
 {\bf Abstract}
\end{center}

We consider two classes of supersymmetric models with nonuniversal gaugino masses at $M_{\rm GUT}$ in an attempt to resolve the apparent muon $g-2$ anomaly encountered in the Standard Model. We explore two distinct scenarios, one in which all gaugino masses have the same sign at $M_{\rm GUT}$, and a second case with opposite sign gaugino masses. The sfermion masses  in both cases are assumed to be universal at $M_{\rm GUT}$. We exploit the non universality among gaugino masses to realize large mass splitting between the colored and non-colored sfermions. Thus, the sleptons can have masses in the few hundred GeV range, whereas the colored sparticles turn out to be an order of magnitude or so heavier. In both models the resolution of the muon $g-2$ anomaly is compatible, among other things, with a $125-126$ GeV Higgs boson mass and the WMAP dark matter bounds.

\newpage

\renewcommand{\thefootnote}{\arabic{footnote}}
\setcounter{footnote}{0}



\section{\label{ch:introduction}Introduction}

The ATLAS and CMS experiments at the Large Hadron Collider (LHC)  have independently reported the discovery \cite{:2012gk, :2012gu} of a  Standard Model (SM) like Higgs boson resonance of mass $m_h \simeq 125-126$ GeV
 using the combined 7~TeV and 8~TeV data. This discovery is compatible with low scale supersymmetry,  since   the Minimal
Supersymmetric Standard Model (MSSM)  predicts an upper bound of $m_h \lesssim 135$ GeV for the lightest  CP-even Higgs boson  \cite{Carena:2002es}. { Note that there exists a class of SO(10)-based supersymmetric models with third family Yukawa unification \cite{yukawaUn} in which the light CP even Higgs boson mass is predicted to be around 125 GeV \cite{Gogoladze:2011aa}.}  On the other hand  no signals for supersymmetric {particles have shown up at the LHC and the} current lower bounds on the colored sparticle masses, are
\begin{equation}
  m_{\tilde{g}} \gtrsim  1.4~{\rm TeV}~ ({\rm for}~ m_{\tilde{g}}\sim m_{\tilde{q}})~~~ {\rm and}~~~
m_{\tilde{g}}\gtrsim 0.9~{\rm TeV}~ ({\rm for}~ m_{\tilde{g}}\ll
m_{\tilde{q}}) ~\cite{Aad:2012fqa,Chatrchyan:2012jx}).
\label{bound1}
\end{equation}
This has created some skepticism about the naturalness arguments { employed for motivating}  low scale supersymmetry.
Although the sparticle mass bounds in Eq. (\ref{bound1}) are mostly derived for the R-parity conserving constrained MSSM (CMSSM), they are  more or less  applicable for a significant class of low scale supersymmetric models.   {In ref. \cite{Bhattacherjee:2013gr} it  was shown that} there is room in the MSSM parameter space for
the bounds in Eq. (\ref{bound1}) to be relaxed, but it is not a large effect  and the models are  specific.
 {The MSSM can accommodate $m_h \simeq 125 \rm \ GeV$  Higgs boson  mass  but it requires either a very large, ${\cal O} (\mathrm{few}-10)$ TeV,  stop quark mass \cite{Ajaib:2012vc}, or a large soft supersymmetry  breaking (SSB) trilinear $A_t$-term, with a stop quark mass of around a TeV \cite{Djouadi:2005gj}. {It is also interesting to note that a Higgs mass  $m_h \simeq 125 \rm \ GeV$    also yields a lower  bound} on the top quark mass, $m_t \gtrsim 168$ GeV, independently from the values of the SSB parameters   \cite{Gogoladze:2014hca}.}

One of the most popular {assumptions} in low scale supersymmetric models is universal SSB {mass terms $(m_{0})$ at  $M_{GUT}$ for the three generations} of sfermions and masses $(M_{1/2})$ {for the $SU(3)_c \times SU(2)_L \times U(1)_Y$   gauginos} . The main motivation for assuming universal  $m_{0}$ is based on the constraints obtained from flavor-changing neutral currents  processes \cite{Martin:1997ns}. Moreover, the assumption of universal gaugino masses is inspired by the possible realization of a grand unified theory. { { With a stop quark mass of  more than 1 TeV} (in order to achieve a 125 GeV  light CP even Higgs boson), and {with  universal SSB parameters} $M_{1/2}$ and $m_{0}$,  the first and second generation {squark masses lie in the multi-TeV range,} and the {corresponding smuon masses lie}  around the TeV scale.}

On the other hand, the SM prediction for the anomalous magnetic moment of the muon \cite{Hagiwara:2011af}, $a_{\mu}=(g-2)_{\mu}/2$, shows a discrepancy with the experimental results \cite{Bennett:2006fi}, {which is quantified as follows:}
\begin{eqnarray}
\Delta a_{\mu}\equiv a_{\mu}({\rm exp})-a_{\mu}({\rm SM})= (28.6 \pm 8.0) \times 10^{-10}
\label{bound2}
\end{eqnarray}
If supersymmetry {is to resolve  }this discrepancy, {one of the  smuons and bino or wino SSB masses  {need} to be quite light}. Thus, it is hard to simultaneously explain the observed Higgs boson mass {and resolve the muon $g-2$ anomaly with universal  sfermion and gaugino  SSB masses at $M_{\rm GUT}$.}  A way out  is to assume non universality in the gaugino sector at $M_{\rm GUT}$.
It is known that the gauginos provide different contributions to the squark and slepton renormalization group equations (RGEs) \cite{Martin:1997ns}.  {It is possible in this case to obtain colored} sparticles with masses around a few TeV, while the {slepton masses are  around a few hundred GeV},
if we assume that the gluino {SSB mass term $M_{3}$ at $M_{\rm GUT}$ is a few times larger} than the bino  and wino SSB mass terms ($M_{1}$ and $M_{2}$). The parameters $m_0$,   $M_{1}$ and $M_{2}$ can be in {the few hundred  GeV range.}

  To retain gauge coupling unification { in the presence of}   nonuniversal   gaugino    masses at $M_{\rm GUT}$, { one could} employ~\cite{Martin:2009ad} non-singlet $F$-terms, compatible with the underlying GUT.  Nonuniversal gauginos can also be generated  from an $F$-term which is a linear combination of two distinct fields of different dimensions \cite{Martin:2013aha}.   One can also consider two distinct sources for supersymmetry breaking \cite{Anandakrishnan:2013cwa}. {With many distinct} possibilities available for realizing nonuniversal gaugino masses  { while keeping universal sfermion mass at $M_{\rm GUT}$}, we employ three independent masses for the  MSSM  gauginos in our study. {There have been several recent attempts to accommodate  $\Delta a_{\mu}$ in Eq. (\ref{bound2})}
within the MSSM framework assuming specific models  for nonuniversal SSB masses  for gauginos \cite{Akula:2013ioa}.
{In a recent paper \cite{Ajaib:2014ana}, we explored the phenomenology of   nonuniversal SSB gaugino masses and split sfermion families
in {the framework} of third family Yukawa unification \cite{yukawaUn}. {It was shown in \cite{Ajaib:2014ana}  that }
 the resolution of the muon $g-2$ anomaly is compatible,
among other things, with the 125 GeV Higgs boson mass,   the WMAP relic dark matter density and excellent $t$-$b$-$\tau$ Yukawa unification.
In this paper we carry out a more thorough investigation of  nonuniversal SSB gaugino masses {and universal sfermion masses at $M_{\rm GUT}$}  without insisting on Yukawa unification.}

The outline of our paper is as follows.
In section \ref{g-2} we briefly describe the dominant contributions to the muon anomalous magnetic moment {arising} from  low scale supersymmetry.
In Section \ref{constraintsSection} we summarize the scanning procedure and the
experimental constraints applied in our analysis.  We also  present the parameter space that we scan over.
In Section \ref{model 1} we assume nonuniversal gauginos at $M_{\rm GUT}$ with $M_3< 0$, $M_{2}> 0$ and $M_1> 0$.
Section \ref{model 2} is dedicated to the case when same sign nonuniversal gaugino masses are assumed at $M_{\rm GUT}$.
The  conclusion are presented in  Section \ref{conclusions}.

\section{\label{g-2}The Muon Anomalous Magnetic Moment}

The leading  contribution {from low scale supersymmetry  to} the muon anomalous magnetic moment is given by \cite{Moroi:1995yh, Martin:2001st}:

\bea
\label{eqq1}
\Delta a_\mu &=&
\frac{\alpha \, m^2_\mu \, \mu\,M_{2} \tan\beta}{4\pi \sin^2\theta_W \, m_{\tilde{\mu}_{L}}^2}
\left[ \frac{f_{\chi}(M_{2}^2/m_{\tilde{\mu}_{L}}^2)-f_{\chi}(\mu^2/m_{\tilde{\mu}_{L}}^2)}{M_2^2-\mu^2} \right]
\nonumber\\
&+&
\frac{\alpha \, m^2_\mu \, \mu\,M_{1} \tan\beta}{4\pi \cos^2\theta_W \, (m_{\tilde{\mu}_{R}}^2 - m_{\tilde{\mu}_{L}}^2)}
\left[\frac{f_{N}(M^2_1/m_{\tilde{\mu}_{R}}^2)}{m_{\tilde{\mu}_{R}}^2} - \frac{f_{N}(M^2_1/m_{\tilde{\mu}_{L}}^2)}{m_{\tilde{\mu}_{L}}^2}\right] \,.
\eea
{Here  $\alpha$ denotes} the fine-structure constant, $m_\mu$  the muon mass, $\mu$   the bilinear Higgs mixing term and $\tan\beta$ is the ratio of the vacuum expectation values (VEVs) of the MSSM Higgs doublets.  $M_1$ and $M_2$ denote the $U(1)_Y$ and $SU(2)$ gaugino masses respectively,  $\theta_W$  is the weak mixing angle {and $m_{\tilde{\mu}_{L}}$ ($m_{\tilde{\mu}_{R}}$) are left (right) handed smuon {masses}}.
The loop functions are defined as follows:
\bea
f_{\chi}(x) &=& \frac{x^2 - 4x + 3 + 2\ln x}{(1-x)^3}~,\qquad ~f_{\chi}(1)=-2/3, \\
f_{N}(x) &=& \frac{ x^2 -1- 2x\ln x}{(1-x)^3}\,,\qquad\qquad f_{N}(1) = -1/3 \, .
\label{eqq2}
\eea
The first term in Eq. (\ref{eqq1}) stands for the dominant contribution coming from one loop diagram with Higgsinos, while the
second term describes inputs from the bino-smuon loop.
As the Higgsino mass $\mu$ increases, {the} first term decreases in Eq. (\ref{eqq1})  and the second term becomes dominant. {The smuons, on the other  hand, must be light, $\lesssim$ few hundred GeV,} in both cases in order to provide sizeable contribution to the muon $g-2$ calculation. Note that the above formula will not be accurate for very large values of $\mu\tan\beta$, according to the decoupling theorem \cite{Moroi:1995yh, Martin:2001st}.
From Eq. (\ref{eqq1}),  the parameters
\begin{equation}
 M_1, \, M_2, \, \mu,\, \tan\beta,  m_{\tilde{\mu}_{L}}, \, m_{\tilde{\mu}_{R}},
\label{eqq3}
\end{equation}
are particularly relevant for the muon $g-2$  calculation, {and we will quantify the desired parameter space later.}
{Since we assume a universal the trilinear SSB term $A_0$, it follows that $A_\mu < \mu \tan\beta$ and we therefore do not consider the trilinear SSB-term contribution in Eq. 3.}


\section{ Scanning Procedure and Experimental Constraints\label{constraintsSection}}

We employ the ISAJET~7.84 package~\cite{ISAJET}
to perform random scans over the parameter space.
In this package, the weak scale values of gauge and third
generation Yukawa couplings are evolved to
$M_{\rm GUT}$ via the MSSM RGEs
in the $\overline{DR}$ regularization scheme.
We do not strictly enforce the unification condition
$g_3=g_1=g_2$ at $M_{\rm GUT}$, since a few percent deviation
from unification can be assigned to unknown GUT-scale threshold
corrections~\cite{Hisano:1992jj}.
With the boundary conditions given at $M_{\rm GUT}$,
all the SSB parameters, along with the gauge and third family Yukawa couplings,
are evolved back to the weak scale $M_{\rm Z}$.

In evaluating  the Yukawa couplings the SUSY threshold
corrections~\cite{Pierce:1996zz} are taken into account
at a common scale  $M_S= \sqrt{m_{\tilde t_L}m_{\tilde t_R}}$.
The entire parameter set is iteratively run between
$M_{\rm Z}$ and $M_{\rm GUT}$ using the full 2-loop RGEs
until a stable solution is obtained.
To better account for the leading-log corrections, one-loop step-beta
functions are adopted for the gauge and Yukawa couplings, and
the SSB scalar mass parameters $m_i$ are extracted from RGEs at appropriate scales
$m_i=m_i(m_i)$.The RGE-improved 1-loop effective potential is minimized
at an optimized scale  $M_S$, which effectively
accounts for the leading 2-loop corrections. Full 1-loop radiative
corrections are incorporated for all sparticle masses.

In scanning the parameter space, we employ the Metropolis-Hastings
algorithm as described in \cite{Belanger:2009ti}.   The data points
collected all satisfy
the requirement of radiative electroweak symmetry breaking  (REWSB)   \cite{Ibanez:1982fr},
with the neutralino in each case being the LSP. After collecting the data, we impose
the mass bounds on all the particles \cite{Nakamura:2010zzi} and use the
IsaTools package~\cite{Baer:2002fv}
to implement the various phenomenological constraints. We successively apply the following experimental constraints on the data that
we acquire from  ISAJET~7.84:
\begin{table}[h!]\centering
\begin{tabular}{rlc}
$123~{\rm GeV} \leq  m_h  \leq 127~{\rm GeV}$~~&\cite{:2012gk,:2012gu}&
\\
$ 0.8 \times 10^{-9} \leq BR(B_s \rightarrow \mu^+ \mu^-) $&$ \leq\, 6.2 \times 10^{-9} \;
 (2\sigma)$        &   \cite{:2007kv}      \\
$2.99 \times 10^{-4} \leq BR(b \rightarrow s \gamma) $&$ \leq\, 3.87 \times 10^{-4} \;
 (2\sigma)$ &   \cite{Barberio:2008fa}  \\
$0.15 \leq \frac{BR(B_u\rightarrow
\tau \nu_{\tau})_{\rm MSSM}}{BR(B_u\rightarrow \tau \nu_{\tau})_{\rm SM}}$&$ \leq\, 2.41 \;
(3\sigma)$. &   \cite{Barberio:2008fa}
\end{tabular}\label{table}
\end{table}

\noindent
{We also implement the following  mass bounds on the sparticle masses:}
\begin{table}[h!]\centering
\begin{tabular}{rlc}
 $m_{\tilde{g}} \gtrsim  1.4~{\rm TeV}~ ({\rm for}~ m_{\tilde{g}}\sim m_{\tilde{q}})$ &~\cite{Aad:2012fqa,Chatrchyan:2012jx}\\
 $m_{\tilde{g}}\gtrsim 1~{\rm TeV}~ ({\rm for}~ m_{\tilde{g}}\ll
m_{\tilde{q}})$ &~\cite{Aad:2012fqa,Chatrchyan:2012jx} \\
$M_A \gtrsim 700~{\rm GeV}$~ $({\rm for}$~ $\tan\beta\simeq 48$). & ~\cite{cms-mA}
\end{tabular}\label{table2}
\end{table}

\noindent
Here $m_{\tilde g}$, $m_{\tilde q}$, $M_A$ {respectively} stand for the gluino, {first and second generation} {squarks} and {the} CP odd Higgs boson masses.

\section{Nonuniversal and opposite sign  gaugino masses \label{model 1}}

In this section we discuss the  {scenario  with nonuniversal and opposite sign  gaugino {masses at $M_{\rm GUT}$,   with the sfermion masses assumed to be universal}. We will show that the muon $g-2$ anomaly can be explained in this model.
We  { perform random scans for} following ranges of the parameters:
\begin{eqnarray}
0 \leq  m_{16}  \leq 3\, \rm{TeV} \nonumber  \\
0 \leq  M_{1}  \leq 5\, \rm{TeV} \nonumber  \\
0 \leq  M_{2}  \leq 5\, \rm{TeV} \nonumber  \\
-5 \leq  M_{3}  \leq 0\, \rm{TeV} \nonumber  \\
-3 \leq A_{0}/m_{16}  \leq 3 \nonumber  \\
2 \leq  \tan\beta  \leq 60 \nonumber \\
0 \leq  m_{10}  \leq 5\, \rm{TeV} \nonumber \\
\mu > 0.
\label{parameterRange}
\end{eqnarray}
Here $m_{16}$ is the universal SSB mass parameter for sfermions, and $M_{1}$, $M_{2}$, and $M_{3}$ denote the SSB gaugino masses for $U(1)_{Y}$, $SU(2)_{L}$ and $SU(3)_{c}$ respectively. $A_{0}$ is the SSB trilinear scalar interaction coupling, $\tan\beta$ is the ratio of the MSSM Higgs vacuum expectation values (VEVs), and $m_{10}$ is the SSB mass term for  the MSSM Higgs doublets.

\begin{figure}[]
\label{fig1}
\subfigure{\includegraphics[scale=1]{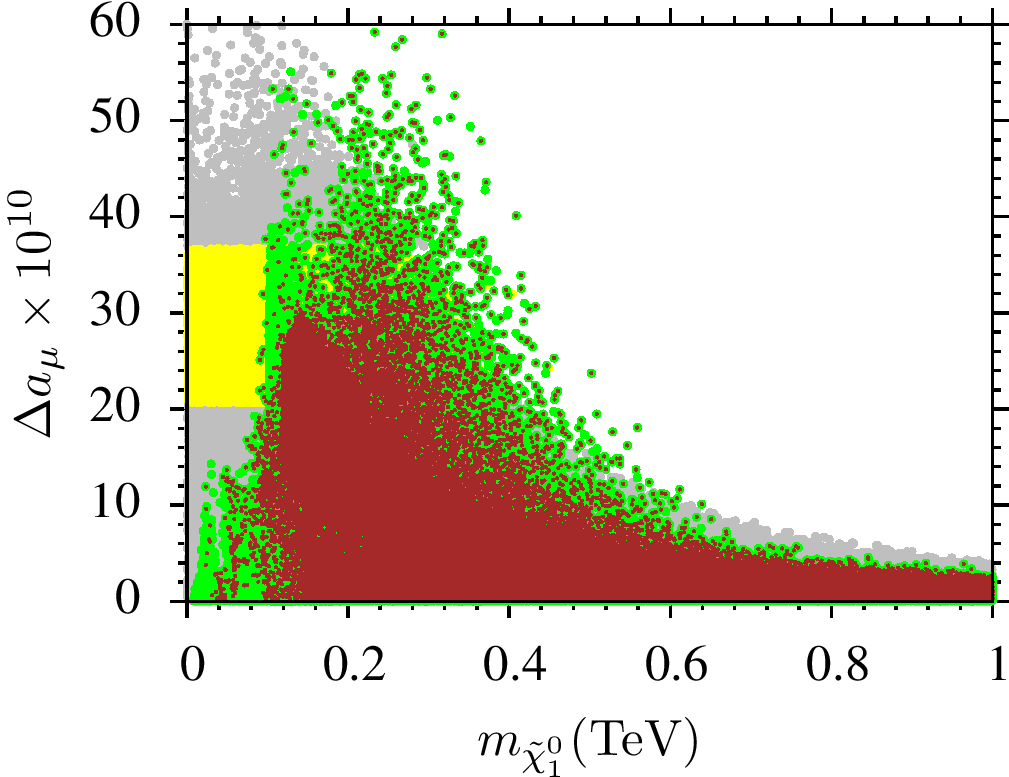}}
\subfigure{\includegraphics[scale=1]{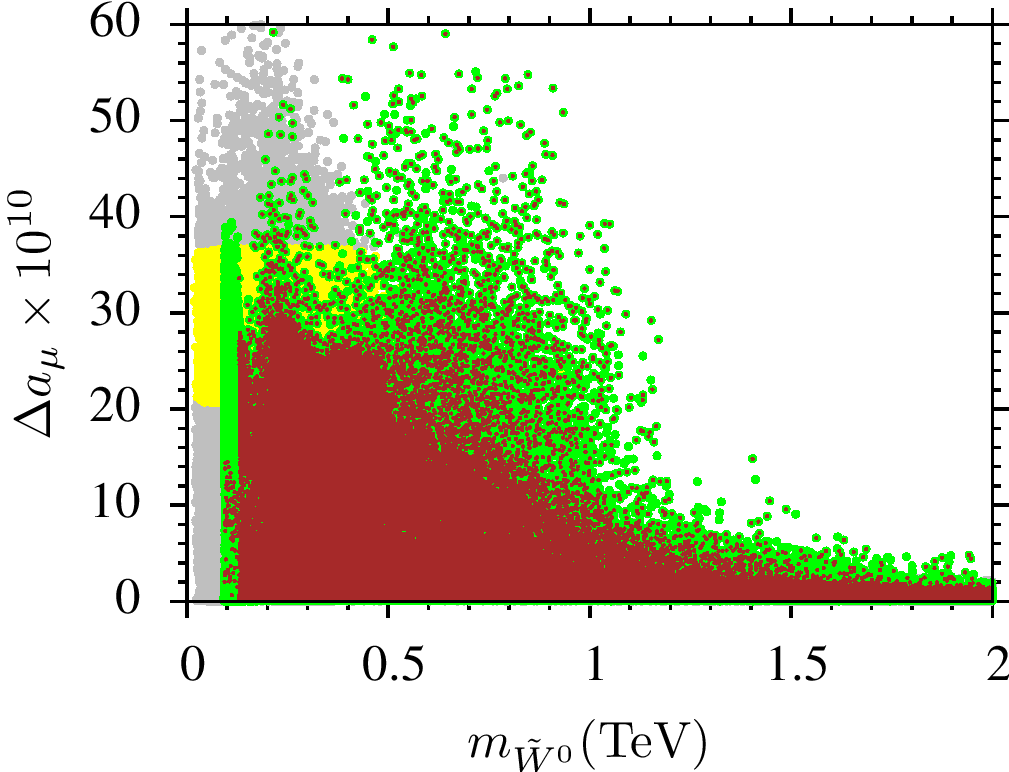}}
\subfigure{\includegraphics[scale=1]{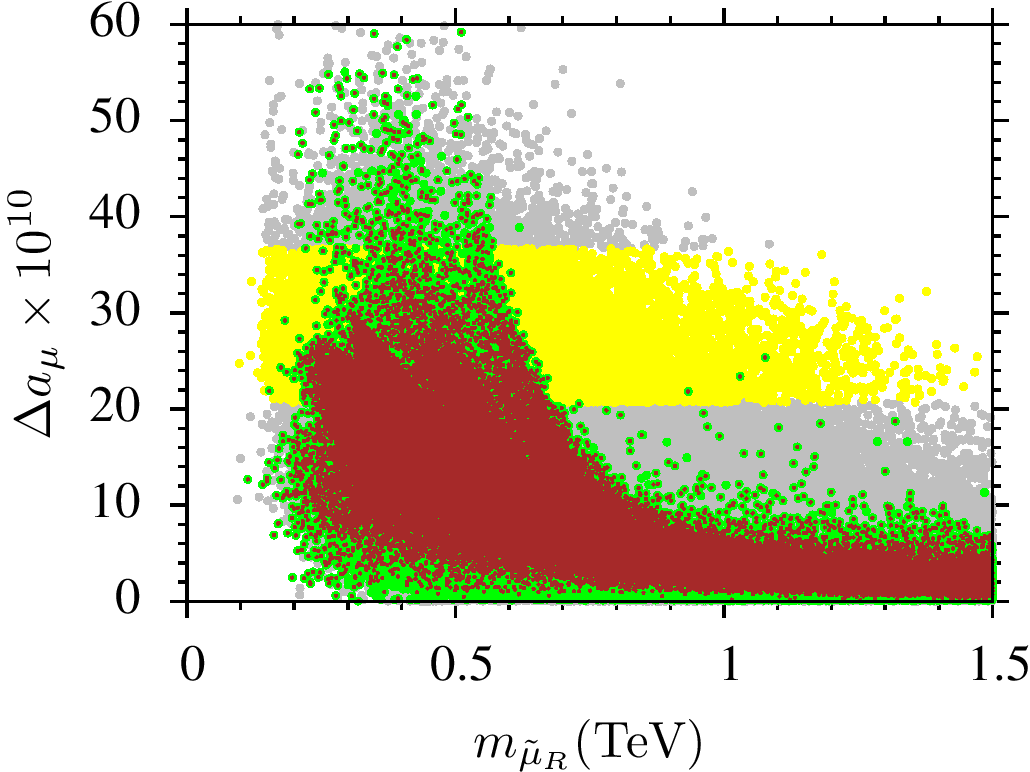}}
\subfigure{\hspace{1.2cm}\includegraphics[scale=1]{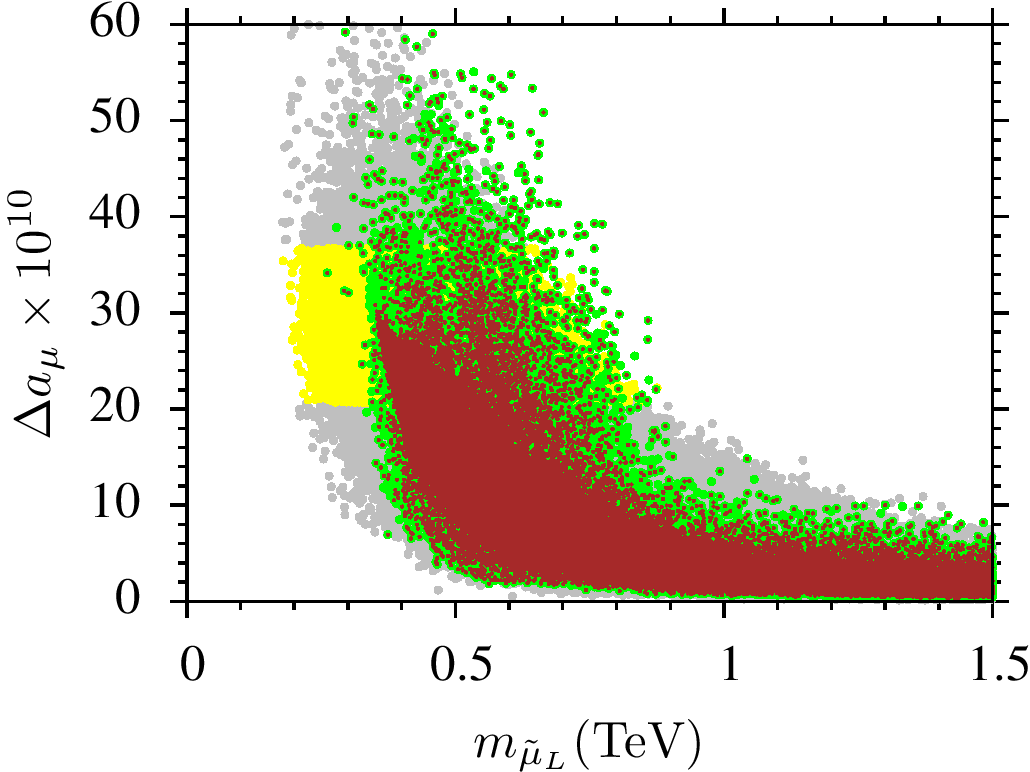}}
\subfigure{\includegraphics[scale=1]{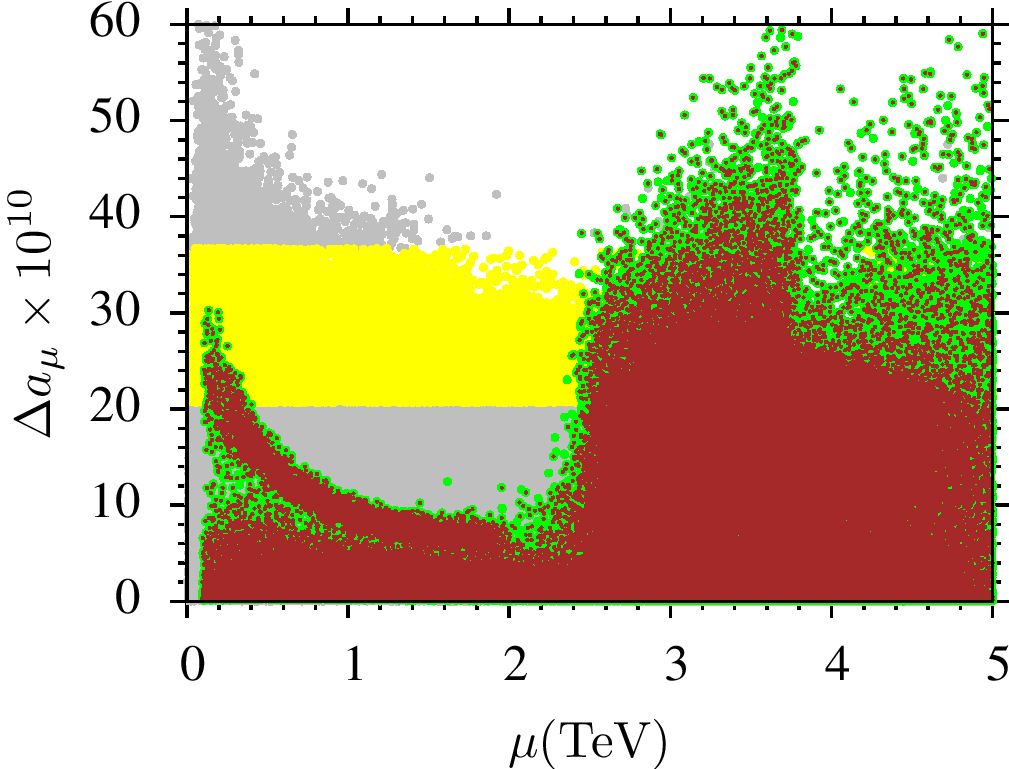}}
\subfigure{\hspace{1.3cm}\includegraphics[scale=1]{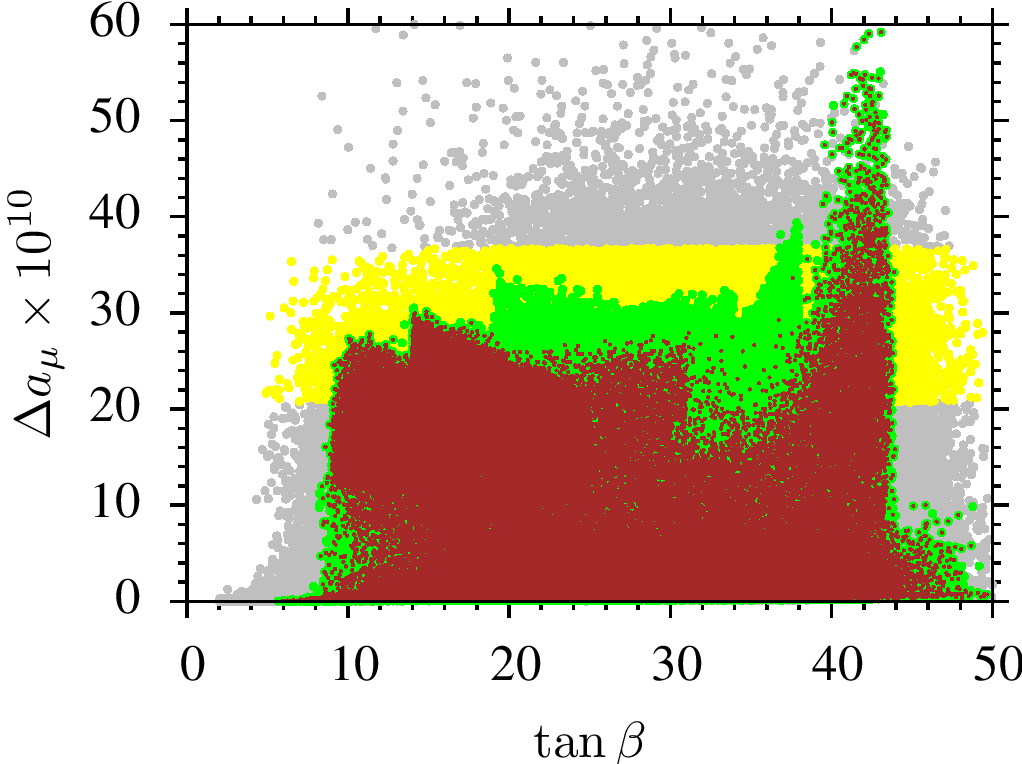}}
\caption{Plots in the $\Delta a_{\mu}-m_{\tilde{\mu}_{R}}$, $\Delta a_{\mu}-m_{\tilde{\mu}_{L}}$, $\Delta a_{\mu}-m_{\tilde{\chi}_{1}^{0}}$, $\Delta a_{\mu}-m_{\tilde{W}^{0}}$,  $\Delta a_{\mu}-\mu$ and  $\Delta a_{\mu}-\tan\beta$ planes. {\it Gray} points are consistent with REWSB  and neutralino LSP.  {\it Yellow} points have $\Delta a_{\mu}$ in {the 1$\sigma$ interval in Eq. (\ref{bound2})}. {\it Green} points form a subset of the {\it gray} {ones}
and satisfy sparticles and Higgs mass bounds and all other constraints described in Section \ref{constraintsSection}.
 {\it Brown} points belong to a   subset of {\it green} and satisfy bound on the LSP neutralino  relic abundance,  $0.001 \leq \Omega h^{2} \leq 1$.  }
\label{fig:1}
\end{figure}

 {As previously mentioned in} Section 2 (Eq. (\ref{eqq3}})),   the  quantities  $M_1$,  $M_2$,  $\mu$, $\tan\beta$,  $m_{\tilde{\mu}_{L}}$,
 $m_{\tilde{\mu}_{R}}$, {play an important role} in the muon $g-2$ calculation. Based on this observation, in Figure 1 we present results in $\Delta a_{\mu}-m_{\tilde{\mu}_{R}}$, $\Delta a_{\mu}-m_{\tilde{\mu}_{L}}$, $\Delta a_{\mu}-m_{\tilde{\chi}_{1}^{0}}$, $\Delta a_{\mu}-\mu$, $\Delta a_{\mu}-\tan\beta$ and $\Delta a_{\mu}-m_{\tilde{W}^{0}}$ planes.
{\it Gray} points are consistent with REWSB  and neutralino LSP.  {\it Yellow} points represent { a subset for which $\Delta a_{\mu}$ lies within the 1$\sigma$ interval in Eq. (\ref{bound2})} . {\it Green} points form a subset of the {\it gray} ones
and satisfy sparticles and Higgs mass bounds and all other constraints described in Section \ref{constraintsSection}.
 {\it Brown} points belong to a   subset of {\it green} and satisfy bound on the LSP neutralino  relic abundance,  $0.001 \leq \Omega h^{2} \leq 1$.
  We have chosen to display our  results for a wider range of $\Omega h^2$ keeping in mind that one can always find points
which are compatible with the current WMAP range for relic abundance \cite{WMAP9} with dedicated scans within the {\it brown} regions.

Overall, from Figure 1 we learn that in order to {provide the desired SUSY contributions} to $\Delta a_\mu$ , while staying consistent with all the experimental constraints described in Section \ref{constraintsSection}, we should impose  the following: $200~{\rm GeV} \lesssim m_{\tilde{\mu}_{R}} \lesssim 700$ GeV,
 $400~{\rm GeV} \lesssim m_{\tilde{\mu}_{L}} \lesssim 800$ GeV,   $100~{\rm GeV} \lesssim m_{\tilde{\chi}_{1}^{0}}\lesssim 400$ GeV, $9 \lesssim \tan\beta \lesssim 44$, $ 100~{\rm GeV} \lesssim m_{\tilde{W}^{0}} \lesssim 1.1$ TeV.

 {The salient features} of the results in Figure \ref{fig:1} can be understood by referring to Eq. (\ref{eqq1}).
We have two dominant contributions at one loop {level arising from sparticles in the loop}. The first term in Eq. (\ref{eqq1}) stands
for contributions involving Higgsinos, while the second term describes the bino-smuon contribution.
As the Higgs bilinear $\mu$ term  increases, the  contribution from the loop involving the Higgsinos decreases, while the
  bino-smuon loop  becomes more relevant. This is the reason why in the spectrum we can have relatively heavy wino, $O(TeV)$, and still maintain sufficient contribution to muon $g-2$.   Since in our setup the gauginos are arbitrary at $M_{\rm GUT}$  {and $m_0$ is $O({\rm few \ hundred})$  GeV or so}, we can have a large difference between the left and right handed smuon masses from RGE running. This allows one to provide a significant  contribution to muon $g-2$  from the loop involving either the left or right handed smuons. Thus,  we can have one of them around a TeV, while the {lighter one is $O({\rm few \ hundred})$ GeV}. { Since in our study  $\mu$ values up to 5 TeV are allowed, the parameter $\tan\beta$ can lie in the fairly wide interval   $9 \lesssim \tan\beta \lesssim 44$.}

\begin{figure}[]
\subfigure{\includegraphics[scale=1]{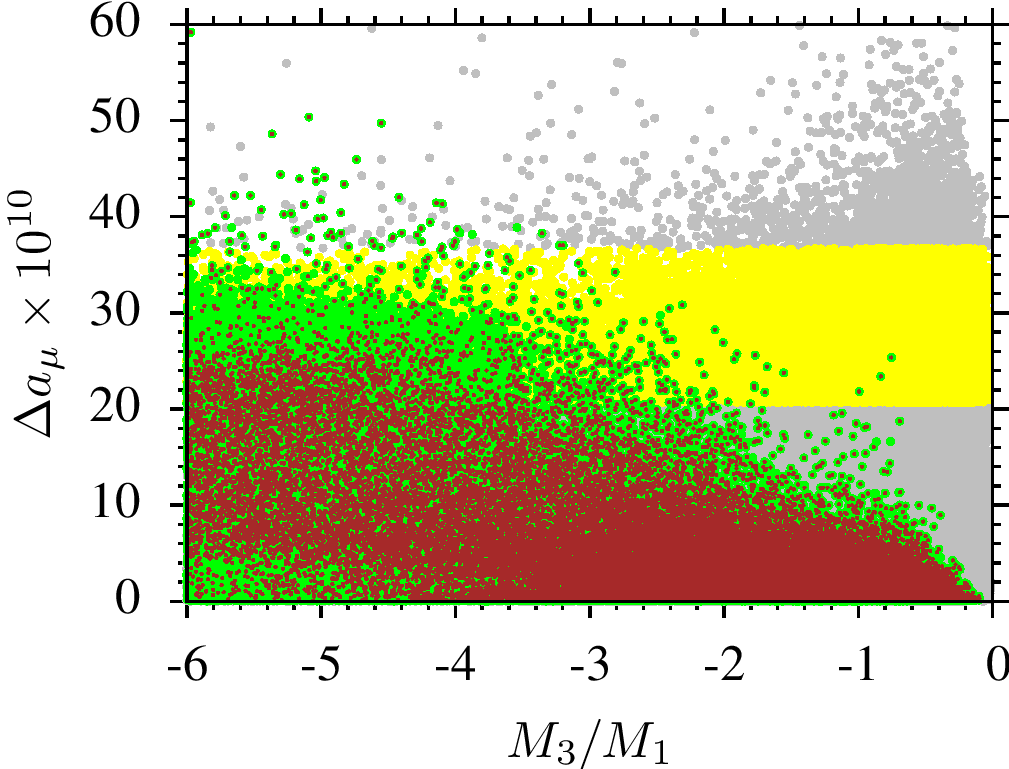}}
\subfigure{\includegraphics[scale=1]{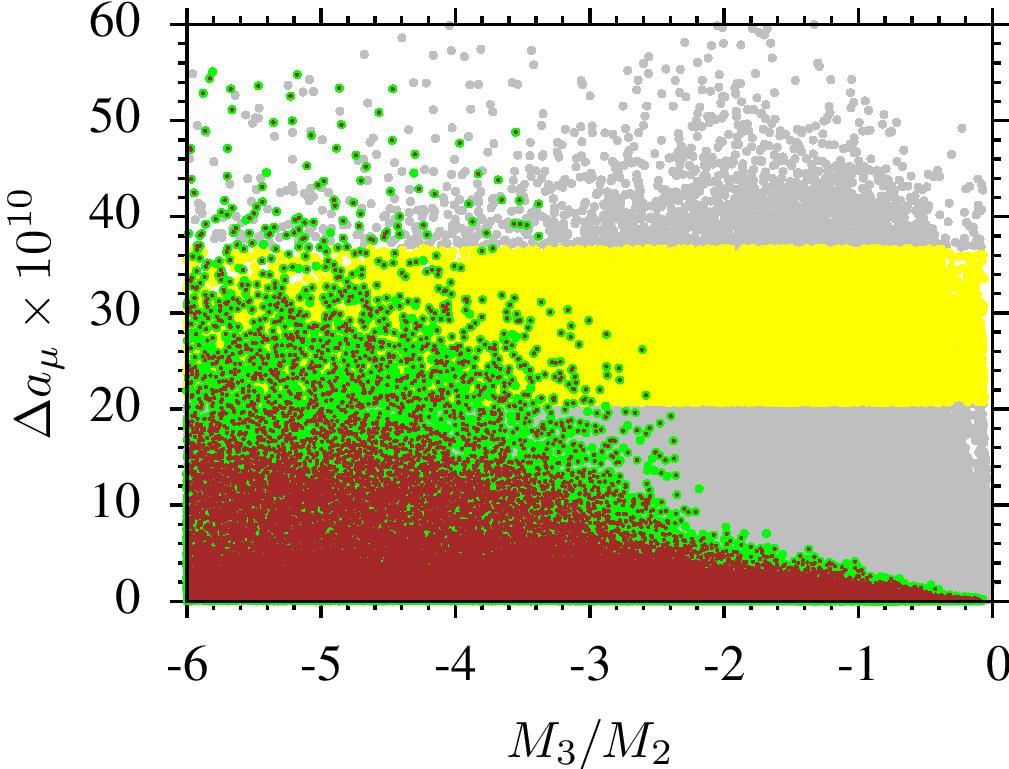}}
\subfigure{\includegraphics[scale=1]{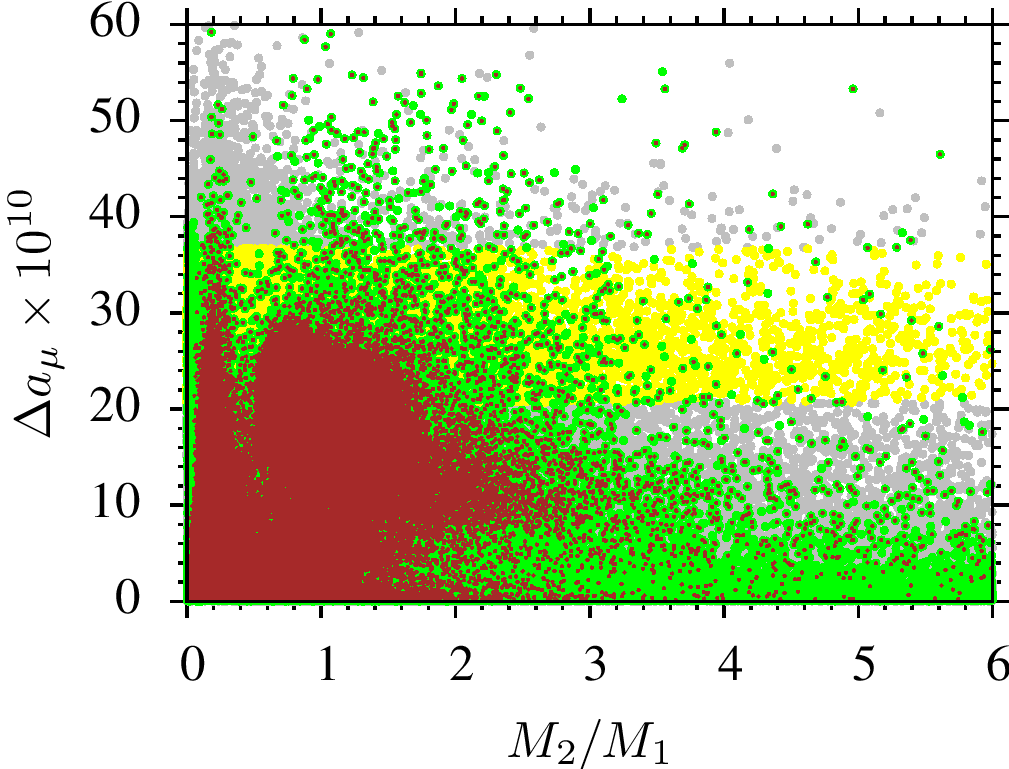}}
\subfigure{\includegraphics[scale=1]{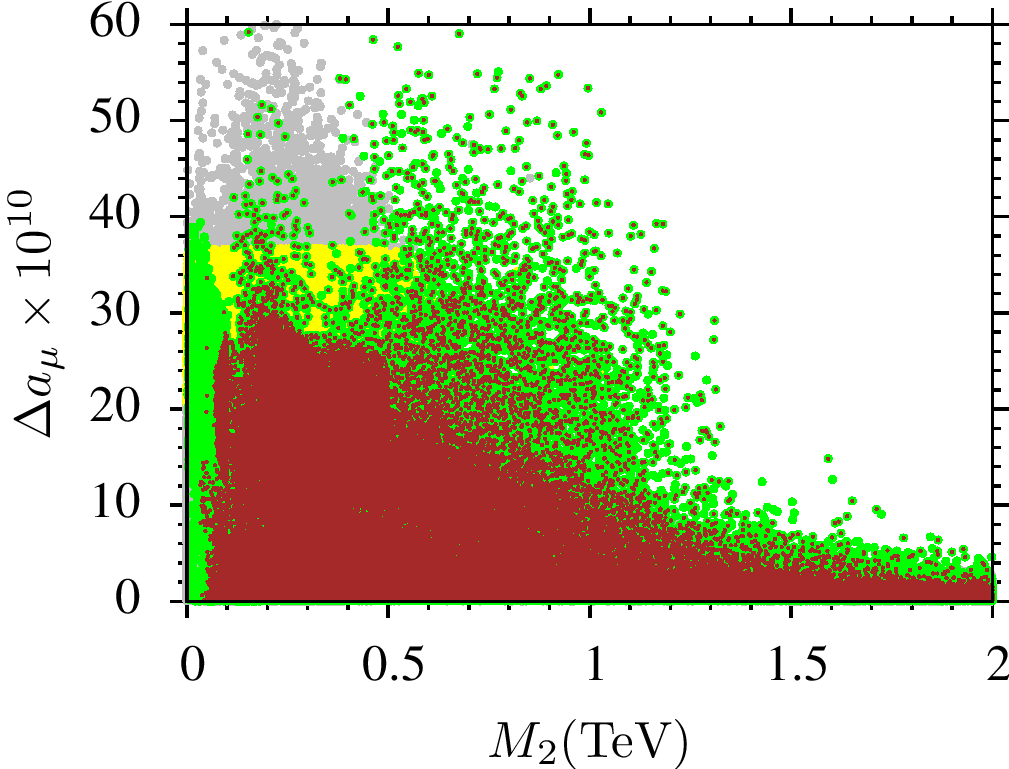}}
\subfigure{\includegraphics[scale=1]{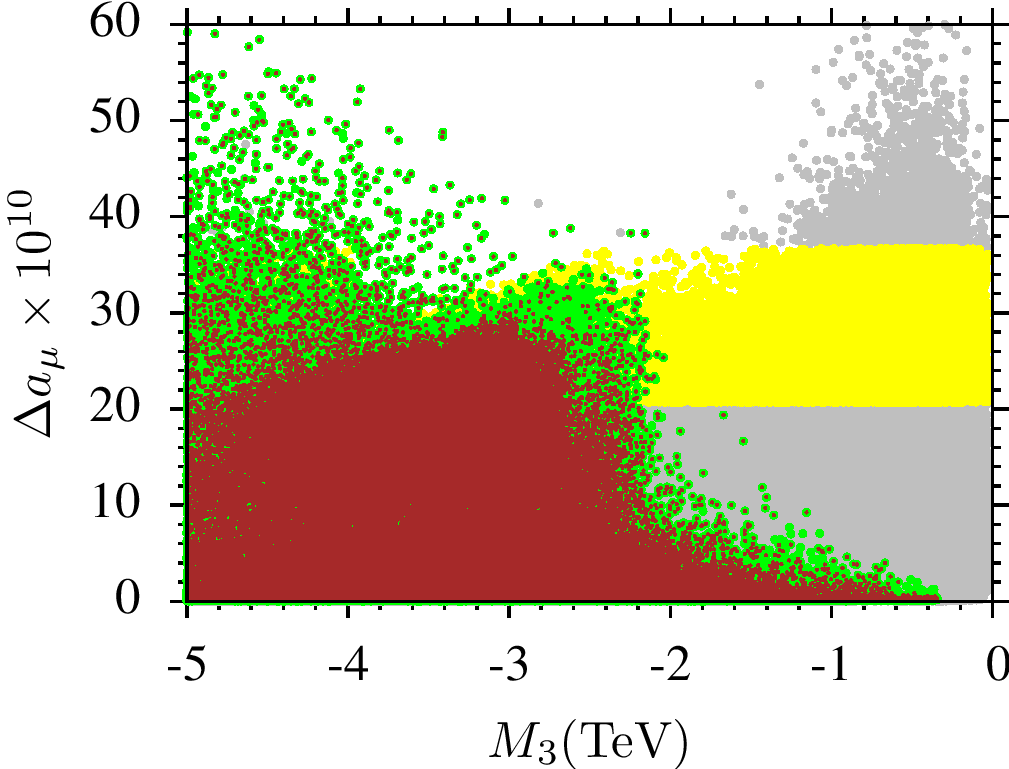}}
\subfigure{\hspace{1.2cm}\includegraphics[scale=1]{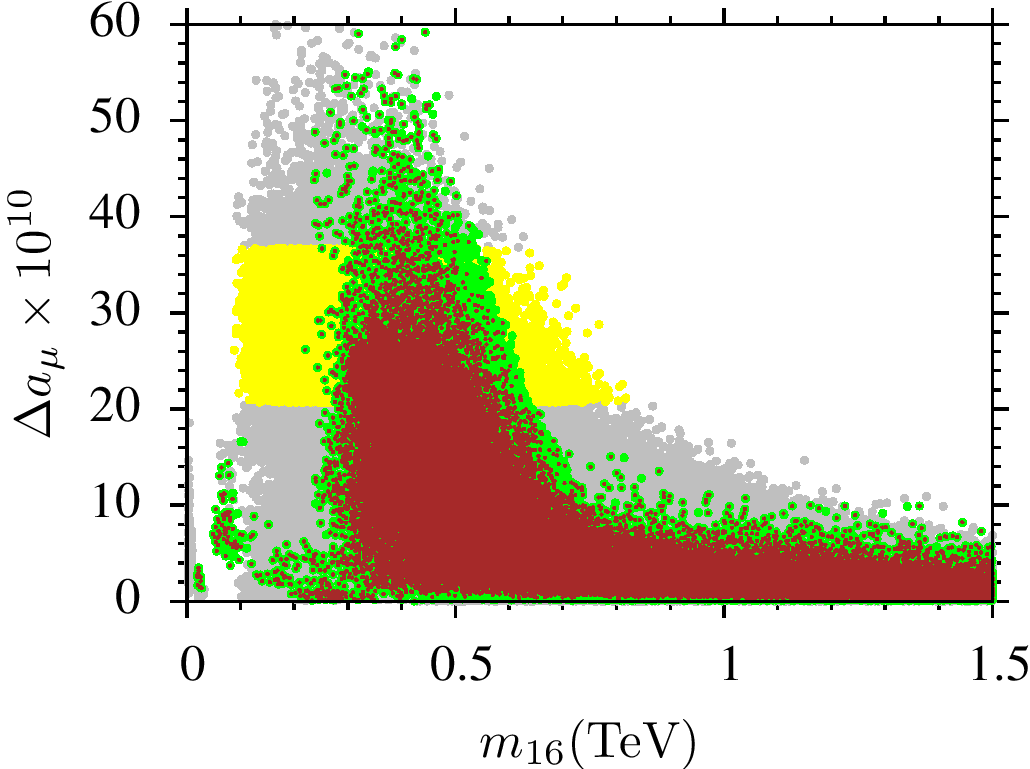}}
\caption{Plots in the $\Delta a_{\mu}-M_{3}/M_{1}$, $\Delta a_{\mu}-M_{3}/M_{2}$, $\Delta a_{\mu}-M_{2}/M_{1}$, $\Delta a_{\mu}-M_{2}$, $\Delta a_{\mu}-M_{3}$ and $\Delta a_{\mu}-m_{16}$ planes. Color coding is the same as in Figure \ref{fig:1}.}
\label{fig:2}
\end{figure}

{The impact of  the muon} $g-2$ anomaly on the fundamental parameters  is presented in Figure \ref{fig:2}, which shows the results in the $\Delta a_{\mu}-M_{3}/M_{1}$, $\Delta a_{\mu}-M_{3}/M_{2}$, $\Delta a_{\mu}-M_{2}/M_{1}$, $\Delta a_{\mu}-M_{2}$, $\Delta a_{\mu}-M_{3}$ and $\Delta a_{\mu}-m_{16}$ planes, with the color coding  the same as in Figure \ref{fig:1}.
{From these results we find the requirements}, $| M_{3}/M_{1} | \leq 0.8$ and $| M_{3}/M_{2} | \leq 2.4$. {The latter ratio is almost the inverse of what was obtained in resolving} the  little hierarchy problem in the MSSM \cite{Gogoladze:2012yf}.
There is no preferred range for the ratio $M_{2}/M_{1}$, { since   diagrams involving only $M_2$ or $M_{1}$  can provide sufficient contribution to muon $g-2$ \cite{Moroi:1995yh, Martin:2001st}. }

 The  $\Delta a_{\mu}-M_{2}$ plane shows that $M_{2} \lesssim 1.3$ TeV at $M_{{\rm GUT}}$, in contrast to $M_{3}$ for which $|M_{3}| \gtrsim 2$ TeV. The last  ($\Delta a_{\mu}-m_{16}$) panel, in Figure 2 shows that  $m_{16}$ cannot be heavier than $\sim 700$ GeV if we require a significant contribution to muon $g-2$.

\begin{figure}[]
\subfigure{\includegraphics[scale=1]{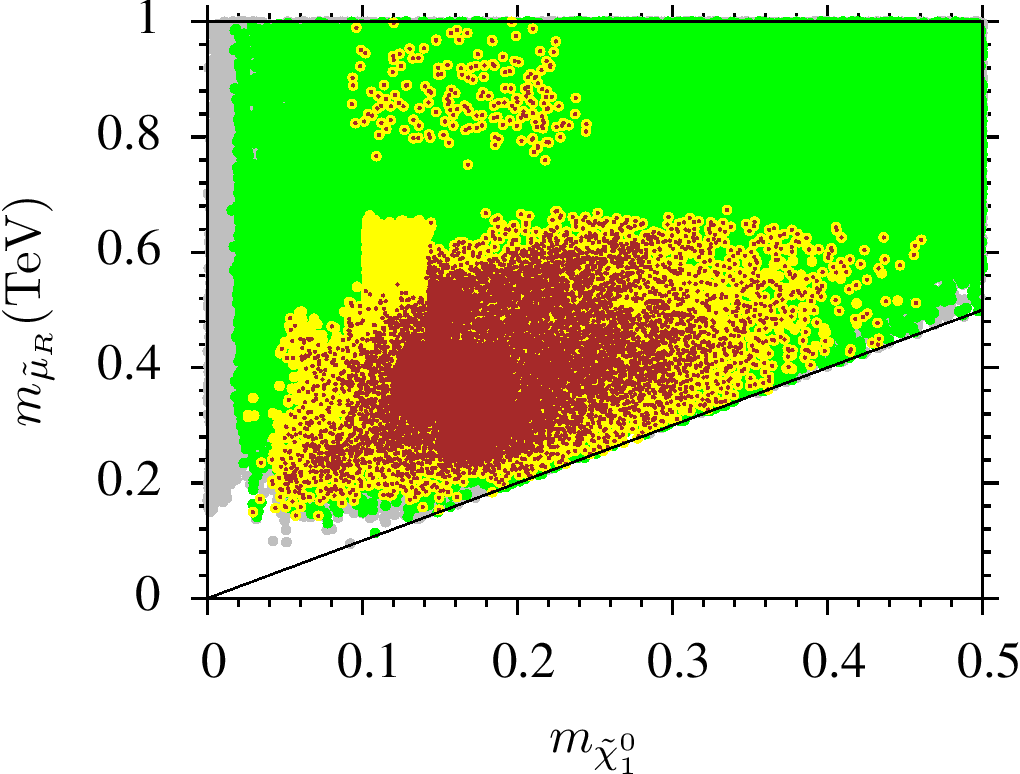}}
\subfigure{\includegraphics[scale=1]{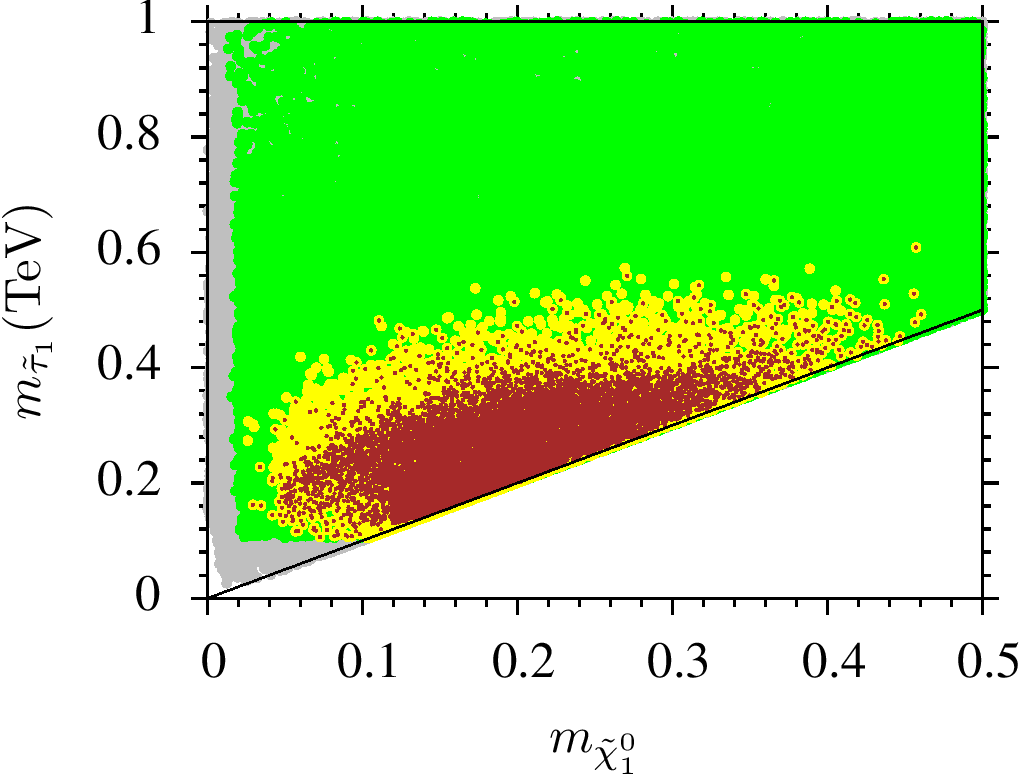}}
\subfigure{\includegraphics[scale=1]{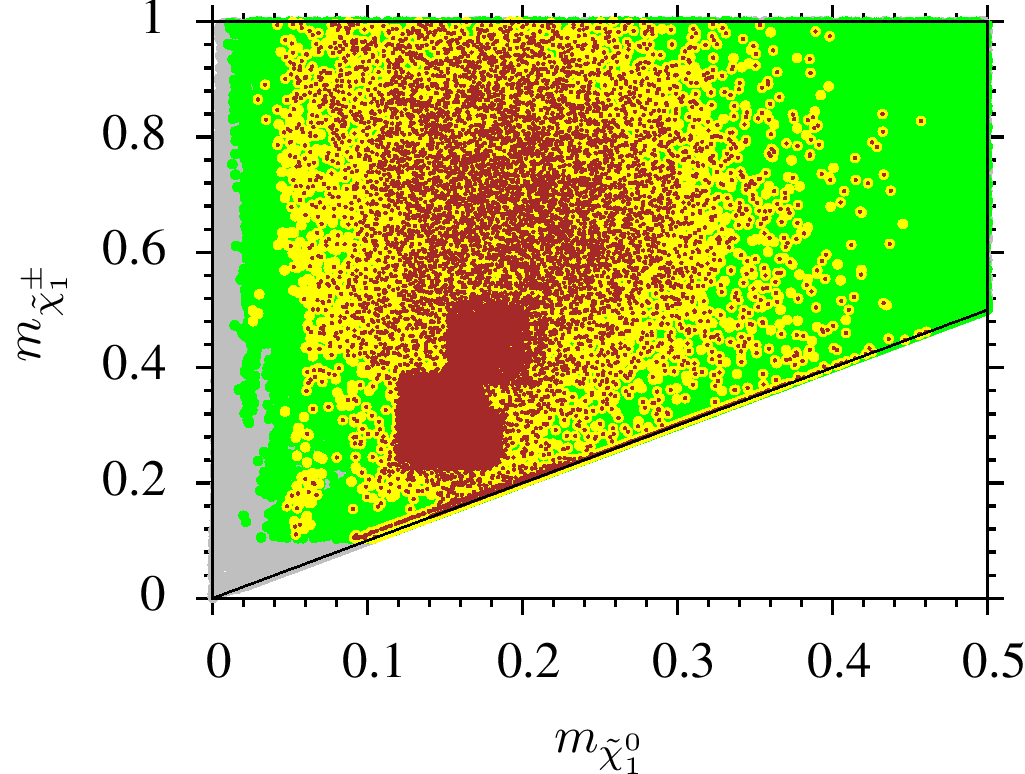}}
\subfigure{\hspace{1.4cm}\includegraphics[scale=1]{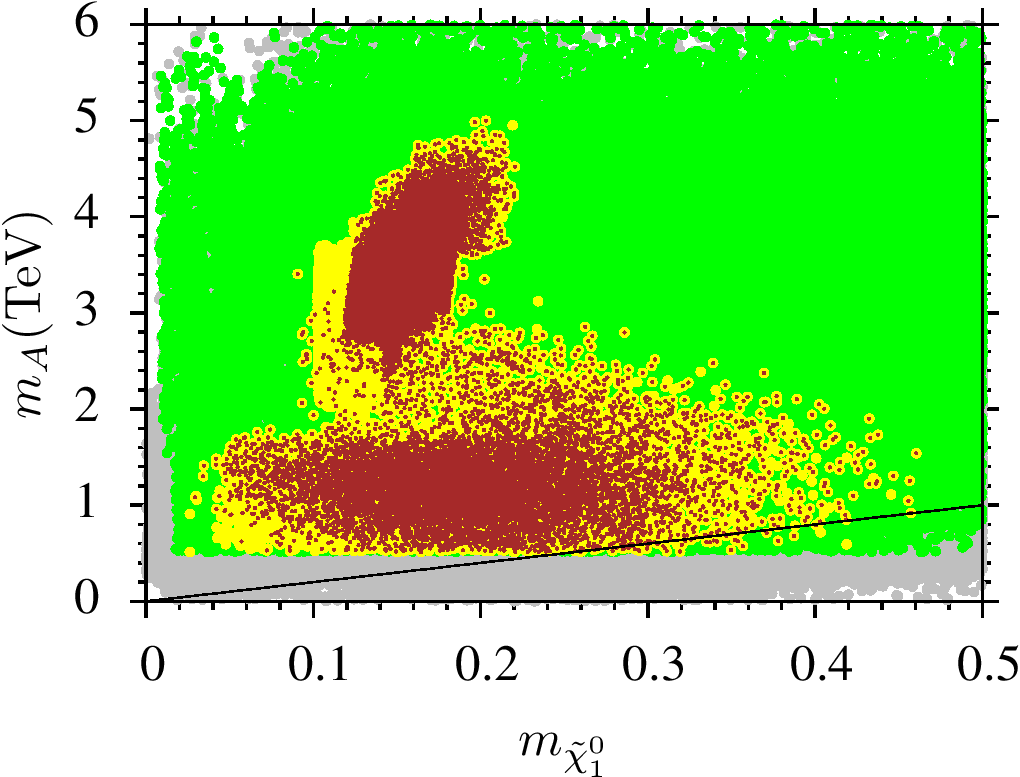}}
\caption{Plots in the $m_{\tilde{\mu}_{R}}-m_{\tilde{\chi}_{1}^{0}}$, $m_{\tilde{\tau}_{1}}-m_{\tilde{\chi}_{1}^{0}}$, $m_{\tilde{\chi}_{1}^{\pm}}-m_{\tilde{\chi}_{1}^{0}}$ and $m_{A}-m_{\tilde{\chi}_{1}^{0}}$ planes. All points are consistent with REWSB and neutralino LSP. Green points satisfy mass bounds and B-physics constraints. Yellow points form a subset of green and they indicate solutions with muon $g-2$ within $1\sigma$ deviation from its theoretical value. Brown points are a subset of yellow and they are consistent with relic abundance of neutralino dark matter in the range $0.001 \leq \Omega h^{2} \leq 1$.}
\label{fig:3}
\end{figure}

The $m_{\tilde{\mu}_{R}}-m_{\tilde{\chi}_{1}^{0}}$, $m_{\tilde{\tau}_{1}}-m_{\tilde{\chi}_{1}^{0}}$, $m_{\tilde{\chi}_{1}^{\pm}}-m_{\tilde{\chi}_{1}^{0}}$ and $m_{A}-m_{\tilde{\chi}_{1}^{0}}$ panels of Figure \ref{fig:3} show that there are  a variety of channels that reduce the relic abundance of neutralino LSP to the desired range applied for the dark matter relic density. All points are consistent with REWSB and neutralino LSP. Green points satisfy all mass bounds and B-physics constraints. Yellow points form a subset of green and they {indicate solutions with the desired contribution} to muon $g-2$. Brown points are a subset of yellow and they are consistent with the relic abundance of LSP neutralino, $0.001 \leq \Omega h^{2} \leq 1$. Since muon $g-2$ requires light smuons, {it is perhaps not surprising to realize the smuon-neutralino} coannihilation scenario, {as  seen in the  $m_{\tilde{\mu}_{R}}-m_{\tilde{\chi}_{1}^{0}}$ plane.} Moreover, in the case of universal sfermion families this  also constrains the third family sfermions to be light. The {lightest stau mass lies} in the range $\sim 100-450$ GeV, and the $m_{\tilde{\tau}_{1}}-m_{\tilde{\chi}_{1}^{0}}$ plane shows the stau-neutralino coannihilation solutions. Similarly the $m_{\tilde{\chi}_{1}^{\pm}}-m_{\tilde{\chi}_{1}^{0}}$ and $m_{A}-m_{\tilde{\chi}_{1}^{0}}$ panels display the chargino-neutralino coannihilation and A-resonance scenarios respectively.

\begin{figure}[]
\begin{center}
\includegraphics[scale=1.]{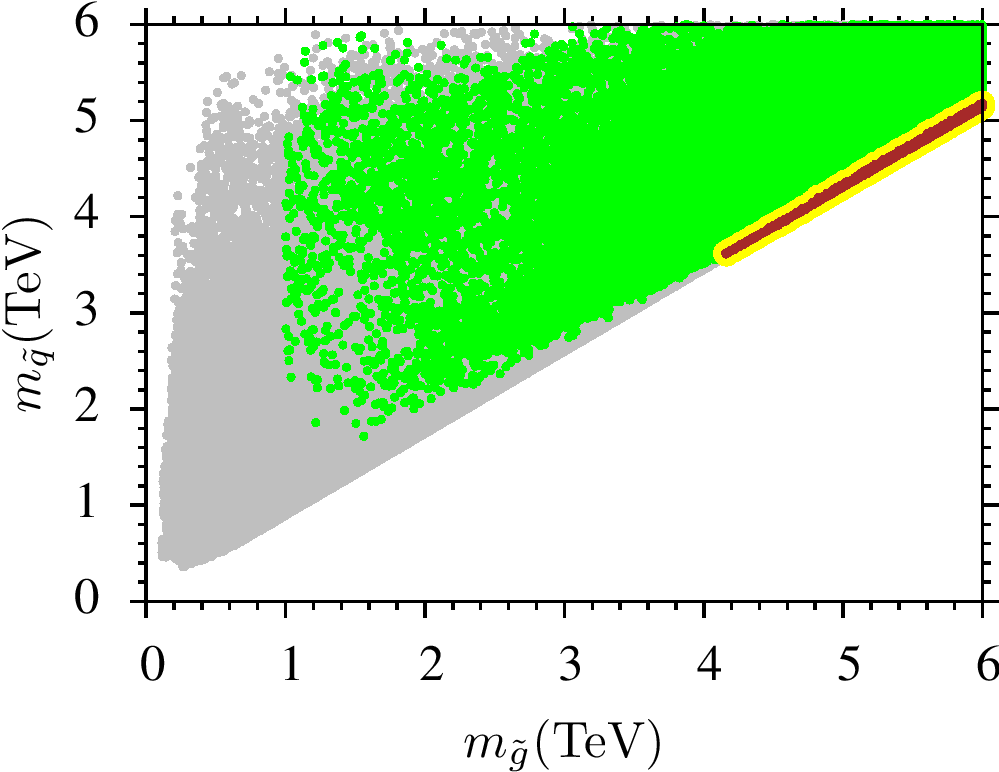}
\end{center}
\caption{Plot in the $m_{\tilde{q}}$-$m_{\tilde{g}}$ plane. Color coding is the same as in Figure \ref{fig:3}.}
\label{fig:4}
\end{figure}

We display the results for the squarks and gluino spectra in the $m_{\tilde{q}}$-$m_{\tilde{g}}$ plane in Figure \ref{fig:4}, with the color coding  the same as in Figure \ref{fig:3}. In this scenario muon $g-2$ allows  solutions with $m_{\tilde{q}}, m_{\tilde{g}} \gtrsim 4$ TeV. Heavy gluino masses are explained with large values of $M_{3}$ at $M_{{\rm GUT}}$ as shown in Figure \ref{fig:2},{ which also  lifts up the squark masses with the resultant} heavy spectrum for squarks, even though the {squarks and sleptons have the  universal} mass  at $M_{{\rm GUT}}$.

Table \ref{tab:1} lists four benchmark points that satisfy the constraints described in Section \ref{constraintsSection} and yield the desired $\Delta a_\mu$.{ For points 1-4, the LSP neutralino relic density} satisfies the WMAP bound, realized via smuon-neutralino, stau-neutralino and chargino-neutralino coannihilation channels and the A-resonance solution, respectively. The gluino is the heaviest colored sparticle for the four benchmark points.

\begin{table}[]\hspace{-1.0cm}
\centering
\begin{tabular}{|c|cccc|}
\hline
\hline
                 & Point 1 & Point 2 & Point 3 & Point 4 \\

\hline
$m_{16}$        & 302.9 & 513.6 & 528 & 422.9 \\
$M_{1}$          & 357 & 428.2 & 554.4  & 618.2 \\
$M_{2} $       & 498.8 & 497.6 & 255.6 & 567.1 \\
$M_{3} $       & -4061 & -4769 & -3027 & -3490 \\
$\tan\beta$      & 9.3 & 39.2 & 42.7 & 43.3 \\
$A_0/m_{16}$      & -0.36 & 0.84 & -0.71 & -0.72 \\
$m_{10}$          & 716.2 & 806.6 & 588.2 & 158.8 \\
$m_t$            & 173.3  & 173.3 & 173.3 & 173.3 \\
\hline
& & & & \\
$\Delta a_{\mu}$  & $ \mathbf{ 24.0\times 10^{-10} } $ & $ \mathbf{28.9\times 10^{-10} } $ & $ \mathbf{ 21.4\times 10^{-10} } $ & $ \mathbf{27.2 \times 10^{-10} } $ \\ & & & & \\

\hline
$m_h$            & \textbf{122.4} & \textbf{124.5} & \textbf{123.0} & \textbf{123.7} \\
$m_H$            & 4278 & 2422 & 1082  & 626.4 \\
$m_A$            & 4250 & 2406 & 1075 & \textbf{622.3} \\
$m_{H^{\pm}}$    & 4279 & 2424 & 1087 & 634.4 \\

\hline
$m_{\tilde{\chi}^0_{1,2}}$
                 & \textbf{188.2},516 & \textbf{225.4}, 528.4 & \textbf{268}, \textbf{287.7} & \textbf{301.6}, 563.3 \\

$m_{\tilde{\chi}^0_{3,4}}$
                 & 41276, 4127 & 4806, 4806 & 3155, 3155 & 3600, 3600 \\

$m_{\tilde{\chi}^{\pm}_{1,2}}$
                 & 520.1, 4089 & 530.4, 4761 & \textbf{289}, 3126 & 565.3, 3567 \\

$m_{\tilde{g}}$  & 8137 & 9471 & 6198 & 7063 \\
\hline $m_{ \tilde{u}_{L,R}}$
                 & 6929, 6947 & 8064, 8088  & 5329, 5348 & 6050, 6059 \\
$m_{\tilde{t}_{1,2}}$
                 & 6004, 6534 & 6998, 7184 & 4617, 4738 & 5244, 5360 \\
\hline $m_{ \tilde{d}_{L,R}}$
                 & 6929, 6952 & 8065, 8088 & 5329, 5350 & 6051, 6062 \\
$m_{\tilde{b}_{1,2}}$
                 & 6494, 6912 & 7122, 7278 & 4642, 4764 & 5234, 5364 \\
\hline
$m_{\tilde{\nu}_{1,2}}$
                 & 267.6 & 491 & 509.5 & 516.9 \\
$m_{\tilde{\nu}_{3}}$
                 &  291.9 & 665.3 & 583.2 & 667.9 \\
\hline
$m_{ \tilde{e}_{L,R}}$
                & 437.1, \textbf{201.2} & 531.1, 502 & 521.9, 556.8 & 526.5, 464.1  \\
$m_{\tilde{\tau}_{1,2}}$
                & \textbf{207.9}, 418.6  & \textbf{262.6}, 840.1 & 302.8, 743.6  & 342.5, 813.4 \\
\hline

$\sigma_{SI}({\rm pb})$
                & $0.26\times 10^{-12}$ & $ 0.56\times 10^{-14} $ & $ 0.34\times 10^{-11} $ & $0.17\times 10^{-10} $ \\

$\sigma_{SD}({\rm pb})$
                & $0.17\times 10^{-9}$ &$ 0.93\times 10^{-10} $ & $ 0.70\times 10^{-9} $ & $0.32\times 10^{-9} $ \\

$\Omega_{CDM}h^{2}$
                &  0.11 & 0.11 & 0.10 & 0.11\\

\hline
\hline
\end{tabular}
\caption{Masses  in this table are in GeV units.
All  points  yield $\Delta a_{\mu}$ in Eq. (\ref{bound2}) within $1\sigma$, and satisfy
 the sparticle mass and B-physics constraints described in Section 3.
 Points 1-4 respectively correspond to smuon-neutralino, stau-neutralino, chargino-neutralino coannihilation channels and A-resonance solutions for neutralino dark mater candidate.
}
\label{tab:1}
\end{table}

\section{Nonuniversal and same sign  gaugino masses \label{model 2}}

\begin{figure}[]
\label{fig5}
\subfigure{\includegraphics[scale=1]{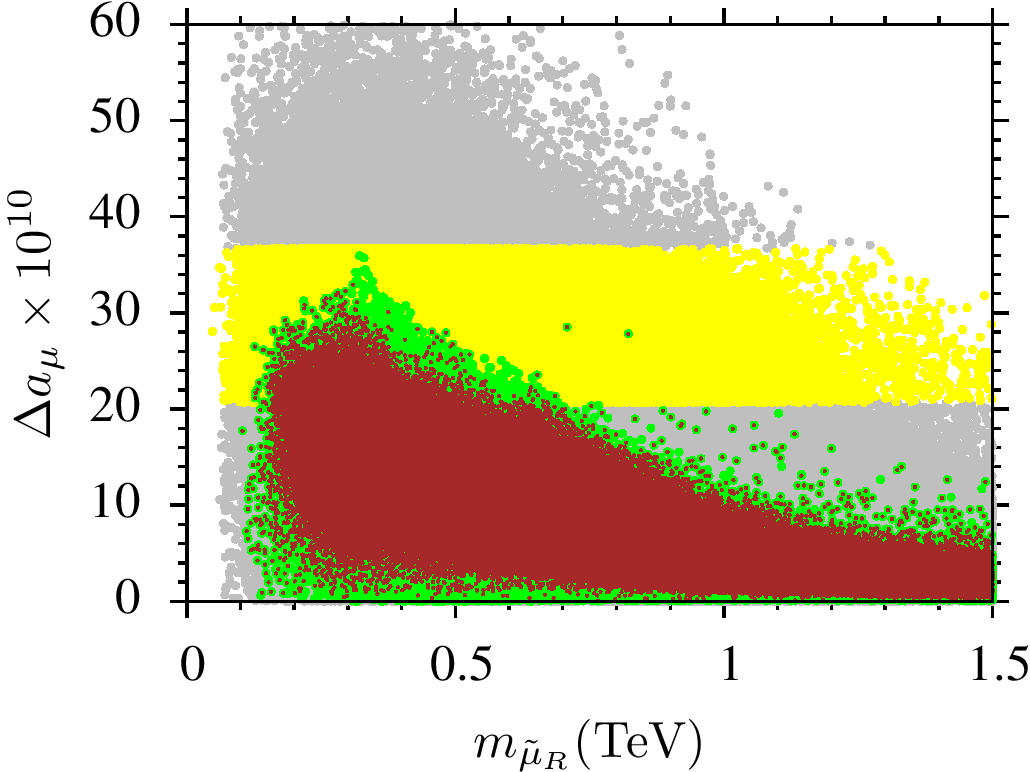}}
\subfigure{\includegraphics[scale=1]{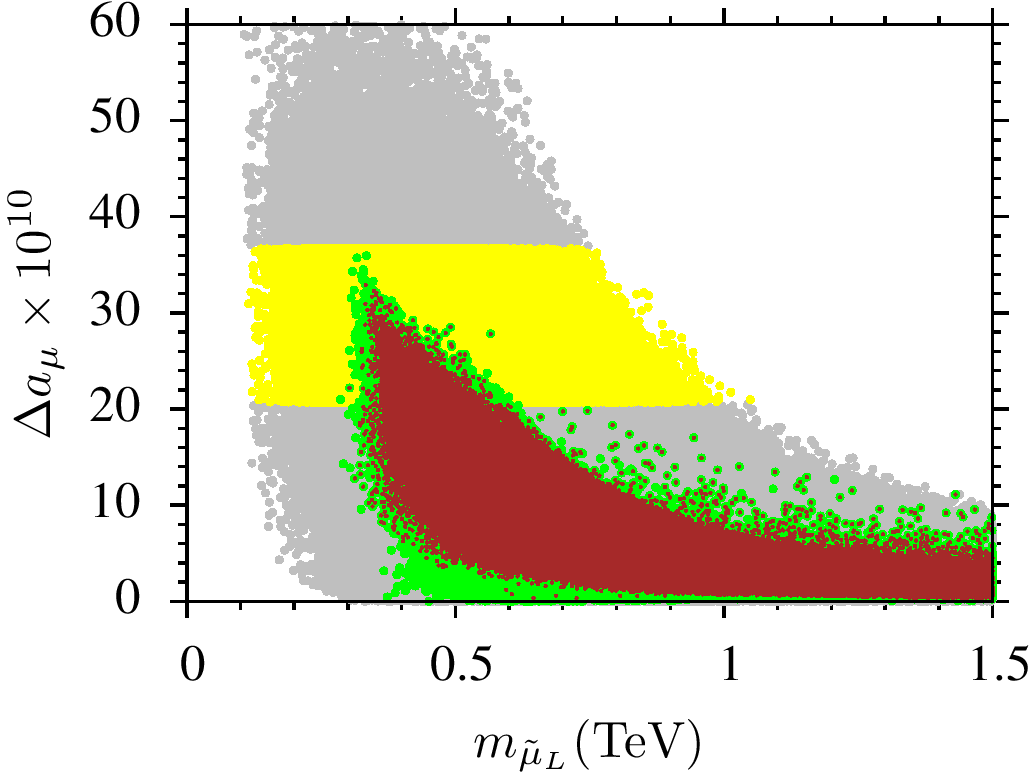}}
\subfigure{\includegraphics[scale=1]{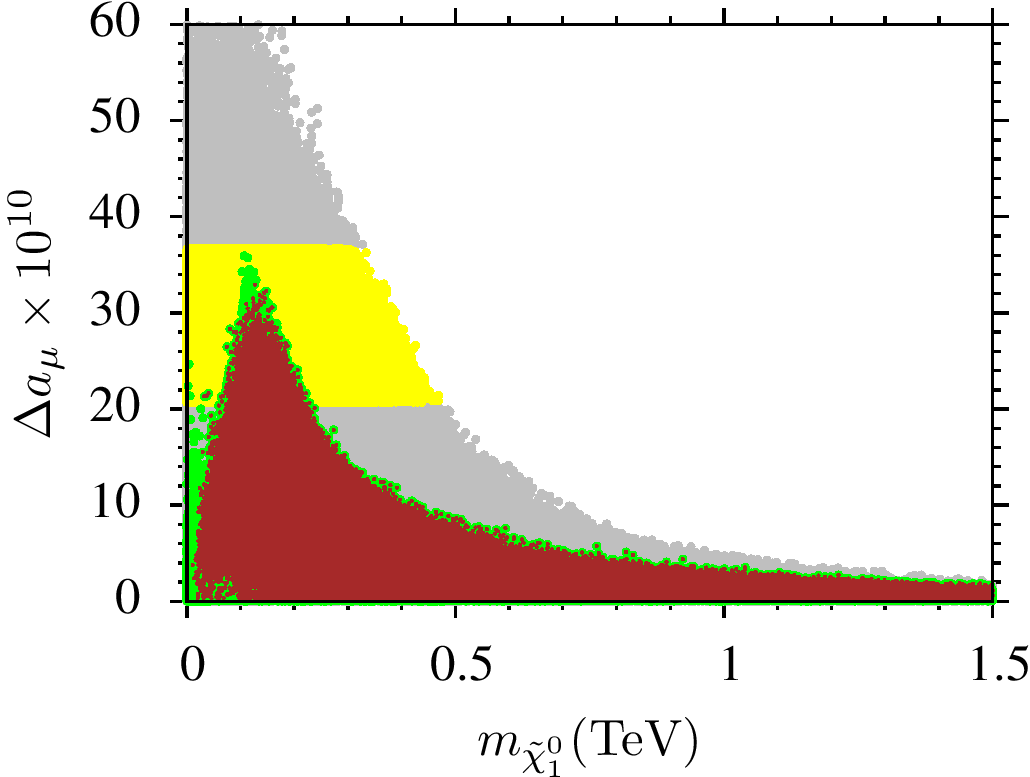}}
\subfigure{\hspace{0.6cm}\includegraphics[scale=1]{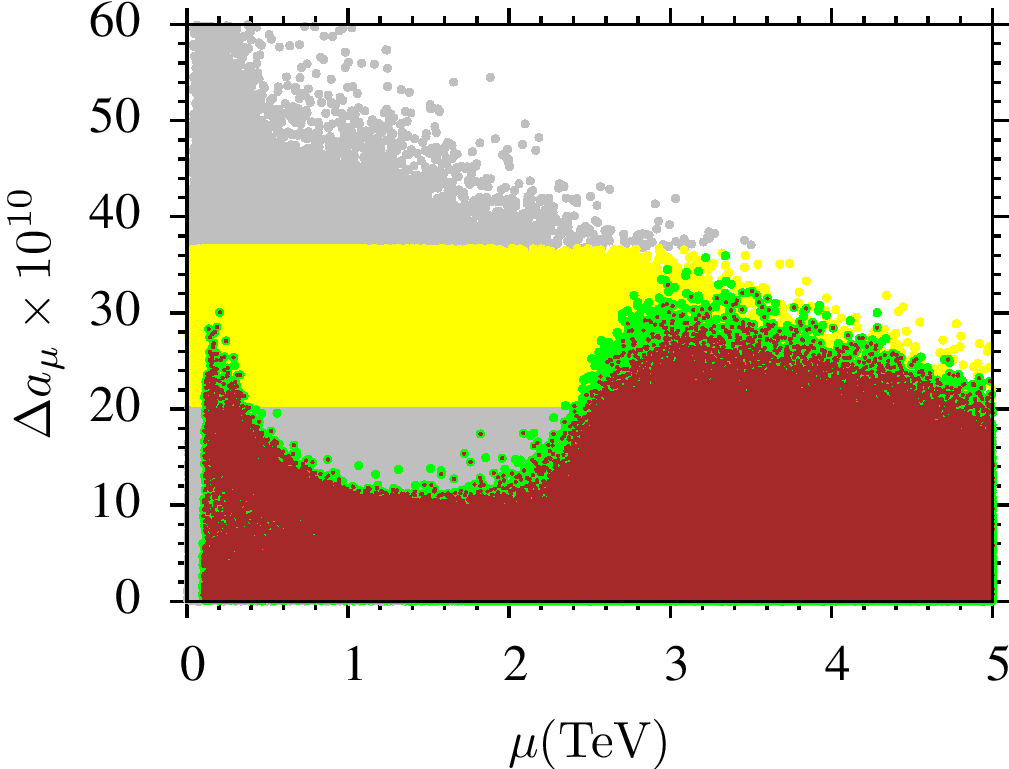}}
\subfigure{\includegraphics[scale=1]{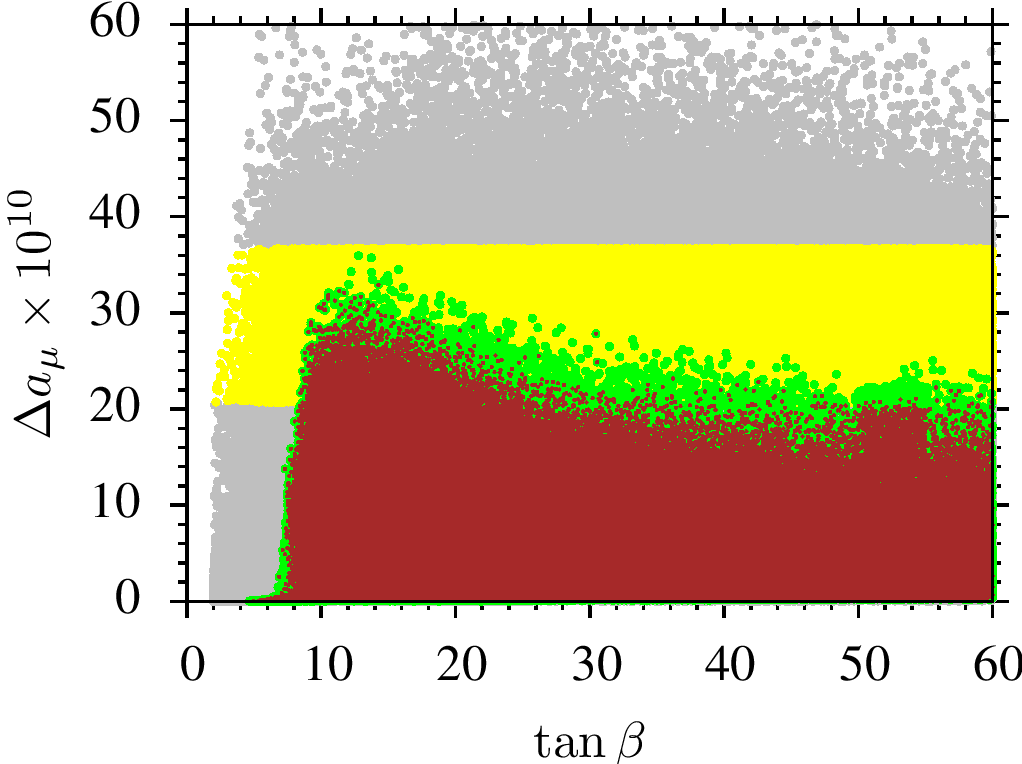}}
\subfigure{\hspace{1.2cm}\includegraphics[scale=1]{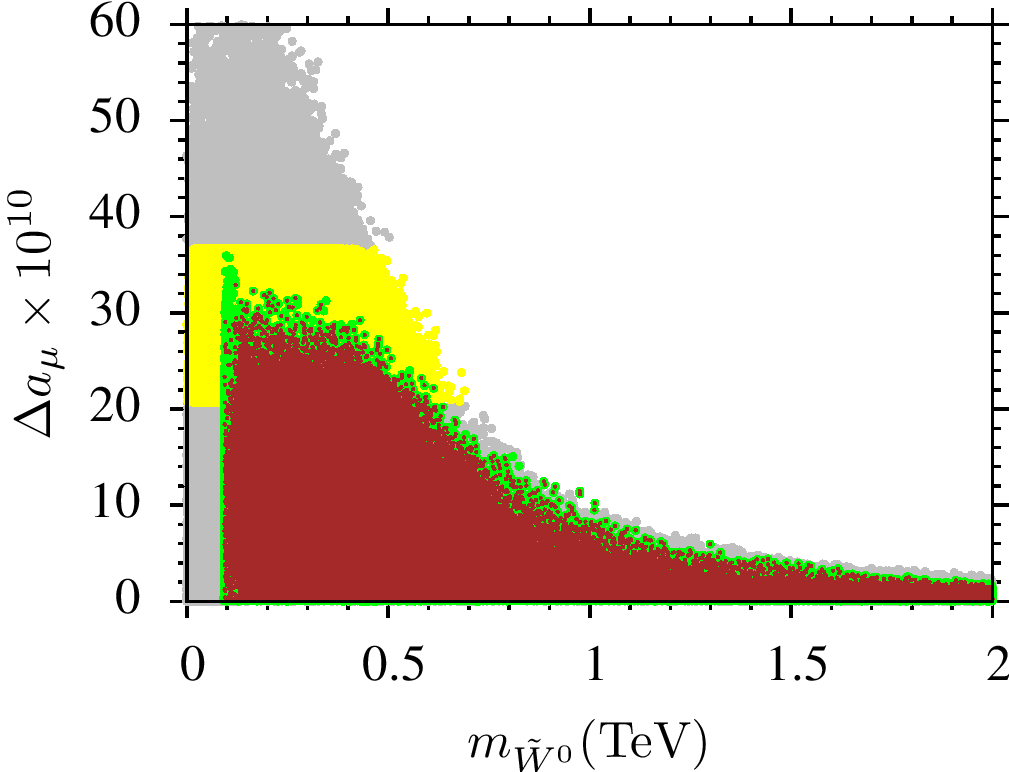}}
\caption{Plots in the $\Delta a_{\mu}-m_{\tilde{\mu}_{R}}$, $\Delta a_{\mu}-m_{\tilde{\mu}_{L}}$, $\Delta a_{\mu}-m_{\tilde{\chi}_{1}^{0}}$, $\Delta a_{\mu}-\mu$, $\Delta a_{\mu}-\tan\beta$ and $\Delta a_{\mu}-m_{\tilde{W}^{0}}$ planes. Color coding is the same as in Figure \ref{fig:1}.}
\end{figure}

In this section we discuss the  scenario  with nonuniversal and same sign  gaugino masses, but with universal sfermion mass  at $M_{\rm GUT}$. The parameter space scanned in this case is as follows:
\begin{align}
0 \leq  m_{16}  \leq 3\, \rm{TeV} \nonumber  \\
0 \leq  M_{1}  \leq 5\, \rm{TeV} \nonumber  \\
0 \leq  M_{2}  \leq 5\, \rm{TeV} \nonumber  \\
0 \leq  M_{3}  \leq 5\, \rm{TeV} \nonumber  \\
-3 \leq A_{0}/m_{16}  \leq 3 \nonumber  \\
2 \leq  tan\beta  \leq 60 \nonumber \\
0 \leq  m_{10}  \leq 5\, \rm{TeV} \nonumber \\
\mu > 0
\label{parameterRange2}
\end{align}
Figure \ref{fig5} shows the results in the $\Delta a_{\mu}-m_{\tilde{\mu}_{R}}$, $\Delta a_{\mu}-m_{\tilde{\mu}_{L}}$, $\Delta a_{\mu}-m_{\tilde{\chi}_{1}^{0}}$, $\Delta a_{\mu}-\mu$, $\Delta a_{\mu}-\tan\beta$ and $\Delta a_{\mu}-m_{\tilde{W}^{0}}$ planes and the color coding is the same as in Figure \ref{fig:1}.
{The results are similar to what we had in the previous section}.  {The small difference arises because of the opposite sign of the gaugino masses, especially when the gluino mass is large compared to the other SSB mass parameters. In this case, the RGE running and supersymmetric thresholds  provide different contributions to the RGEs of the stop quark masses and  $A_t$ \cite{Pierce:1996zz,Martin:1993zk,Gogoladze:2009bd}.  On the other hand, these two quantities provide the dominant contribution to the radiative correction to the mass of the mass of the light CP even Higgs. We find that the reduction of green points in Figure 5 compared to Figure 1 occurs because of the  Higgs boson mass bound, $122~{\rm GeV}\leq m_h \leq 127~\rm{GeV}$.}

{The figure in} $\Delta a_{\mu}-m_{\tilde{\mu}_{R}}$ plane shows that the right-handed smuon can be as heavy as 1 TeV or so, while the left-handed smuon is bounded in {a region of order} $350 - 700$ GeV as seen in the $\Delta a_{\mu}-m_{\tilde{\mu}_{L}}$ plane. We can see from the $\Delta a_{\mu}-m_{\tilde{\chi}_{1}^{0}}$ plane that only  solutions with a light LSP ($\sim 100 - 300$ GeV) are allowed by the muon $g-2$ constraint.   {The $\Delta a_{\mu}-\mu$ plane indicates that a sizable contribution to muon $g-2$ prefers mostly  large values of $\mu$, {but smaller values are} also possible as discussed in the previous section. It is possible to find solutions with a wide range of $\tan\beta$, even though the contributions to $g-2$ slightly decrease as $\tan\beta$ increases. Also, the wino cannot be heavier than $\sim 700$ GeV in order to have significant contributions to muon $g-2$.}

\begin{figure}[]
\subfigure{\includegraphics[scale=1]{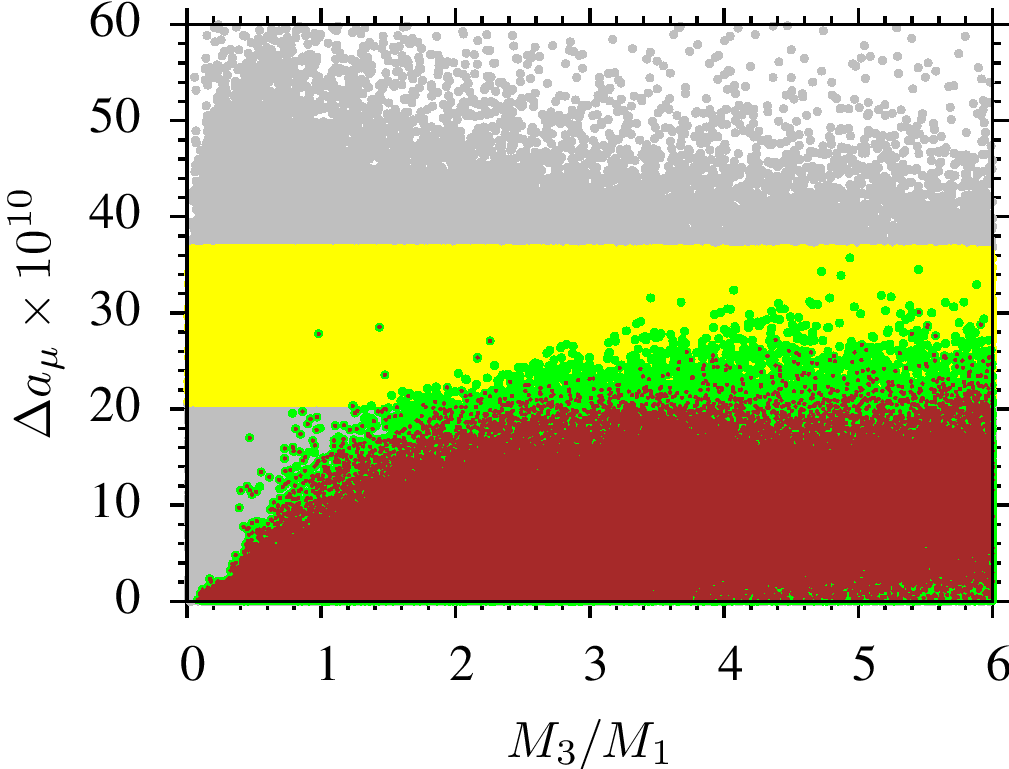}}
\subfigure{\includegraphics[scale=1]{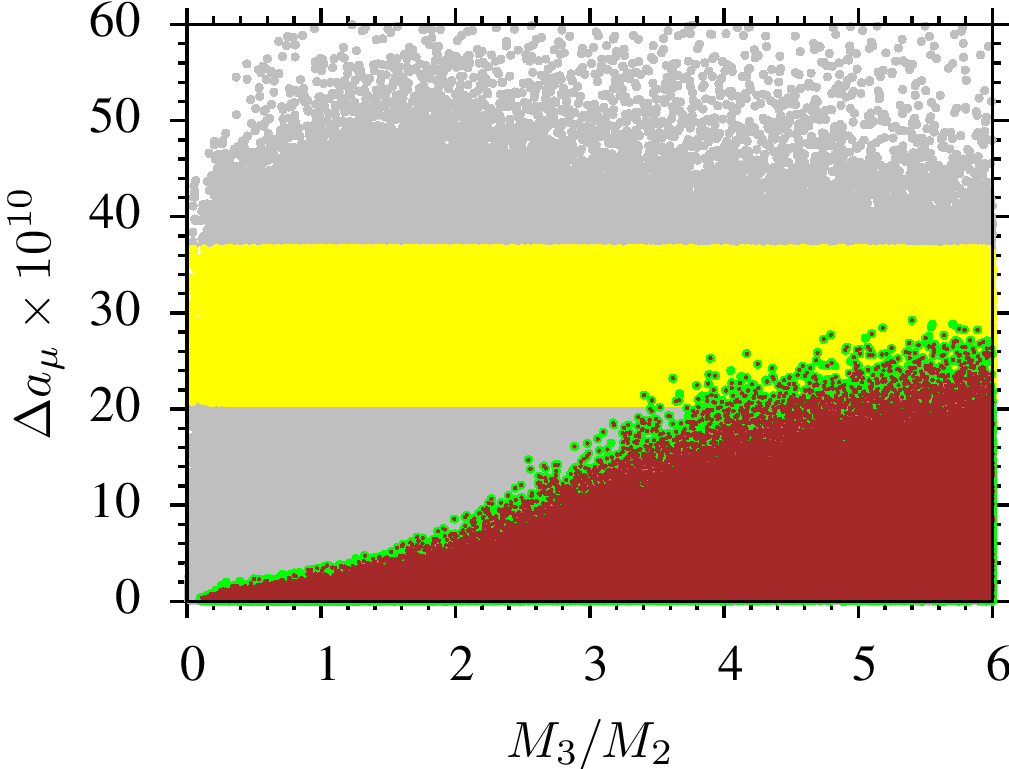}}
\subfigure{\includegraphics[scale=1]{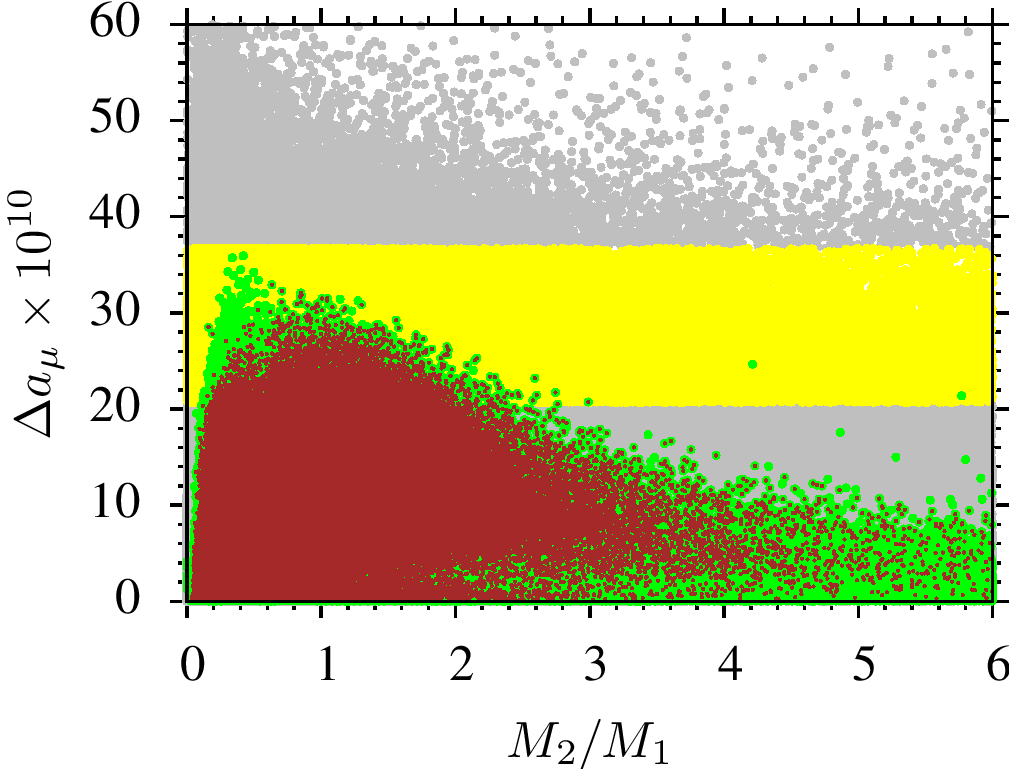}}
\subfigure{\includegraphics[scale=1]{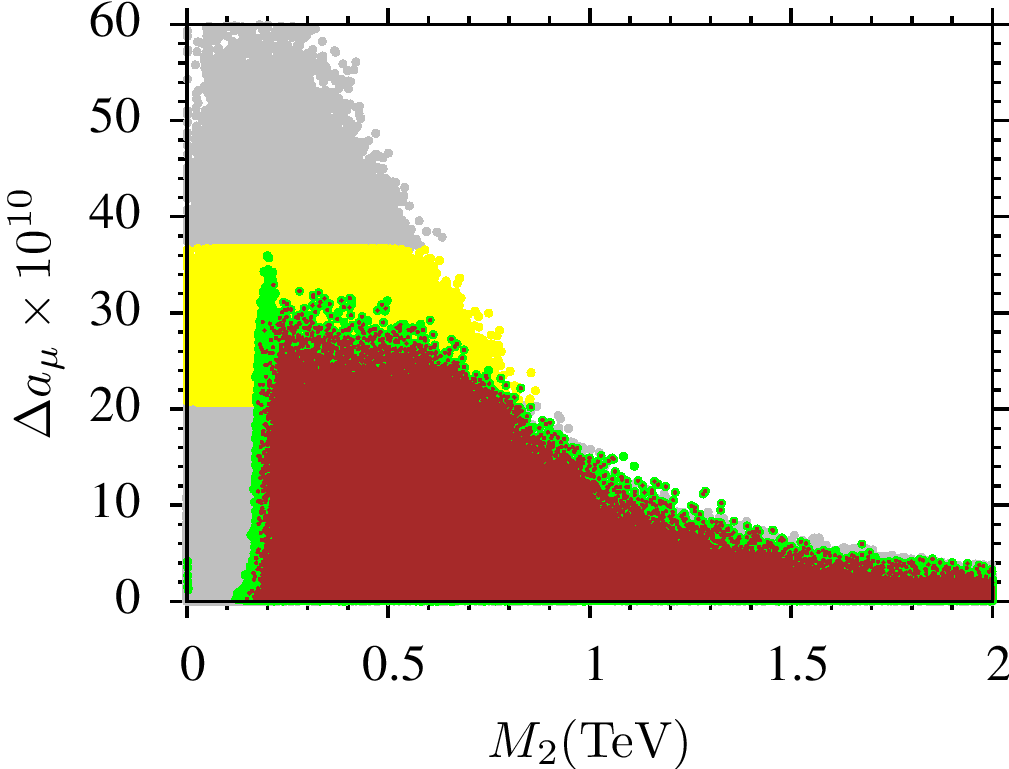}}
\subfigure{\includegraphics[scale=1]{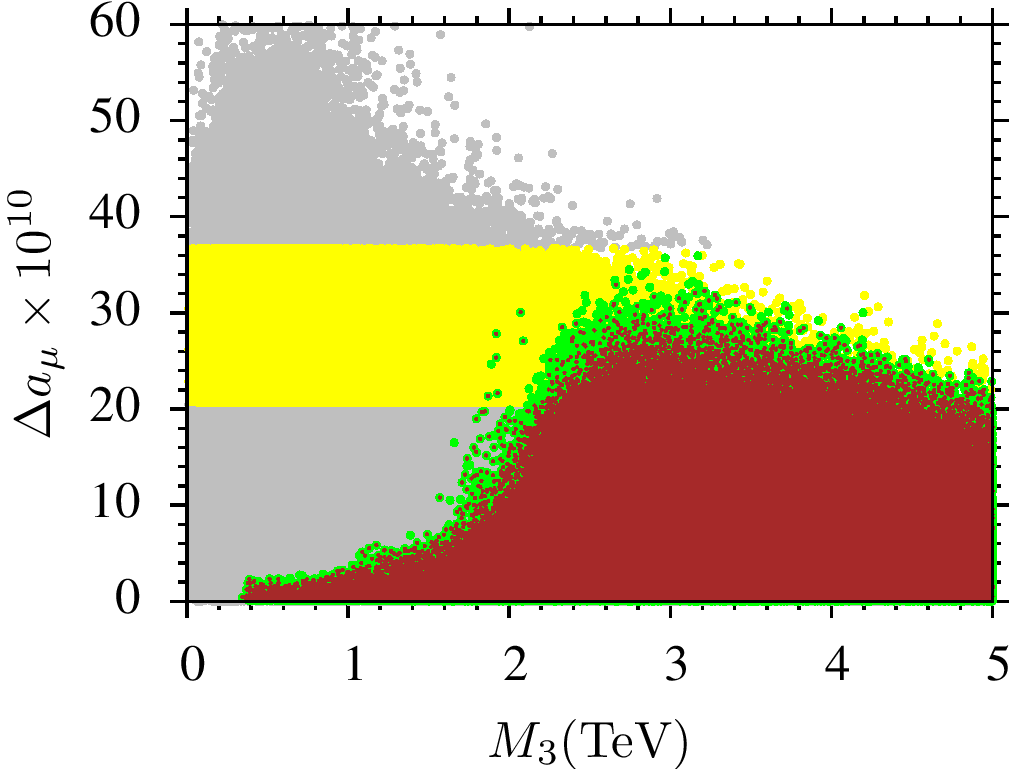}}
\subfigure{\hspace{1.2cm}\includegraphics[scale=1]{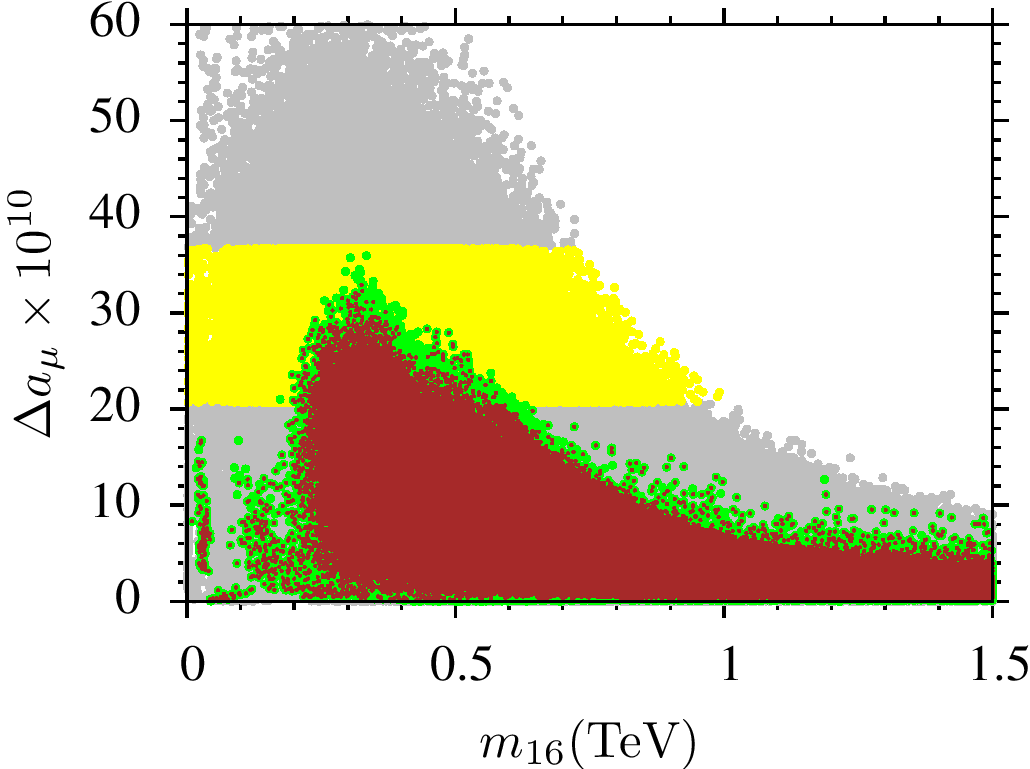}}
\caption{plots in $\Delta a_{\mu}-M_{3}/M_{1}$, $\Delta a_{\mu}-M_{3}/M_{2}$, $\Delta a_{\mu}-M_{2}/M_{1}$, $\Delta a_{\mu}-M_{2}$, $\Delta a_{\mu}-M_{3}$ and $\Delta a_{\mu}-m_{16}$ planes. Color coding is the same as in Figure \ref{fig:1}.}
\label{fig6}
\end{figure}

\begin{figure}[]
\subfigure{\includegraphics[scale=1]{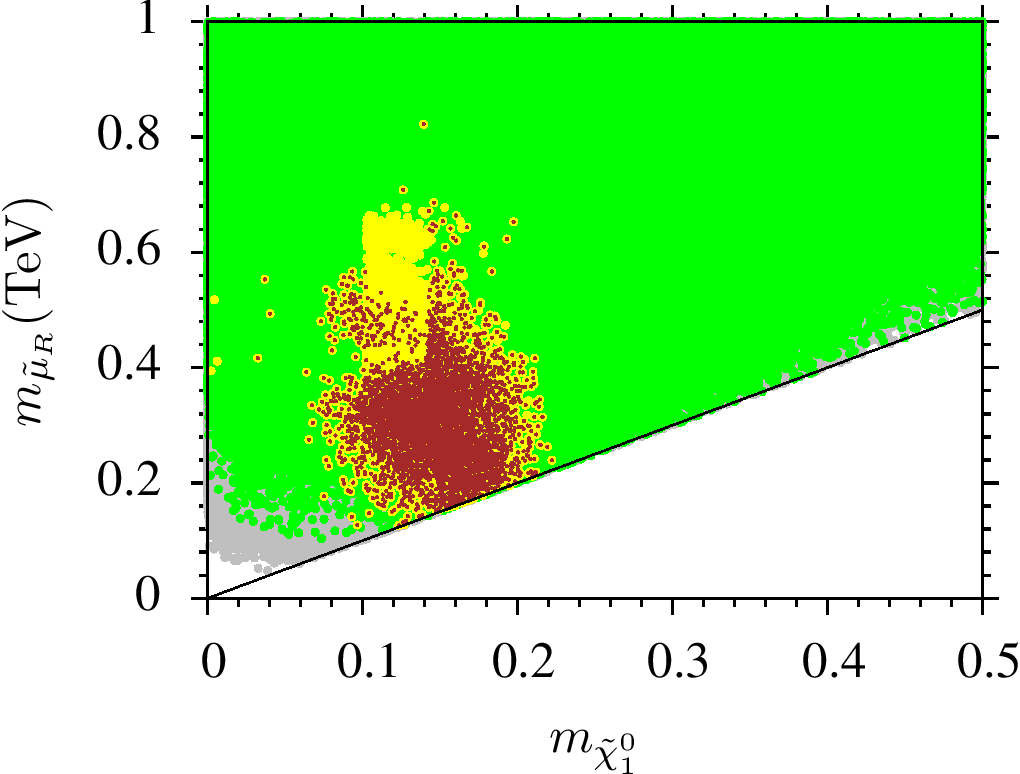}}
\subfigure{\includegraphics[scale=1]{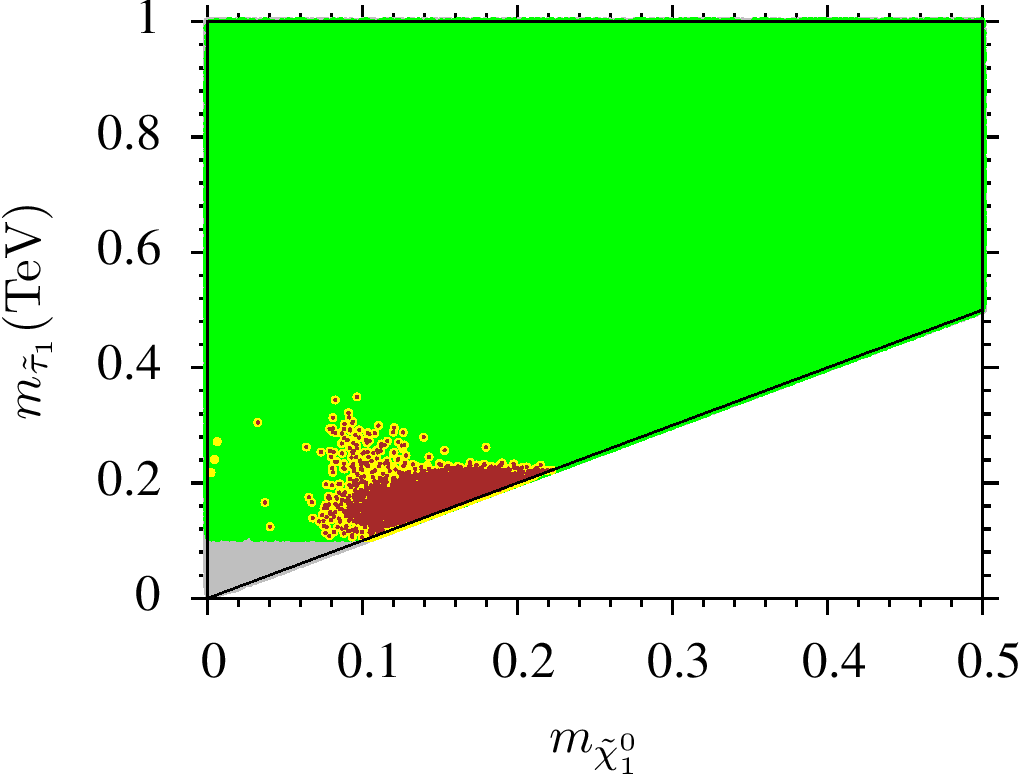}}
\subfigure{\includegraphics[scale=1]{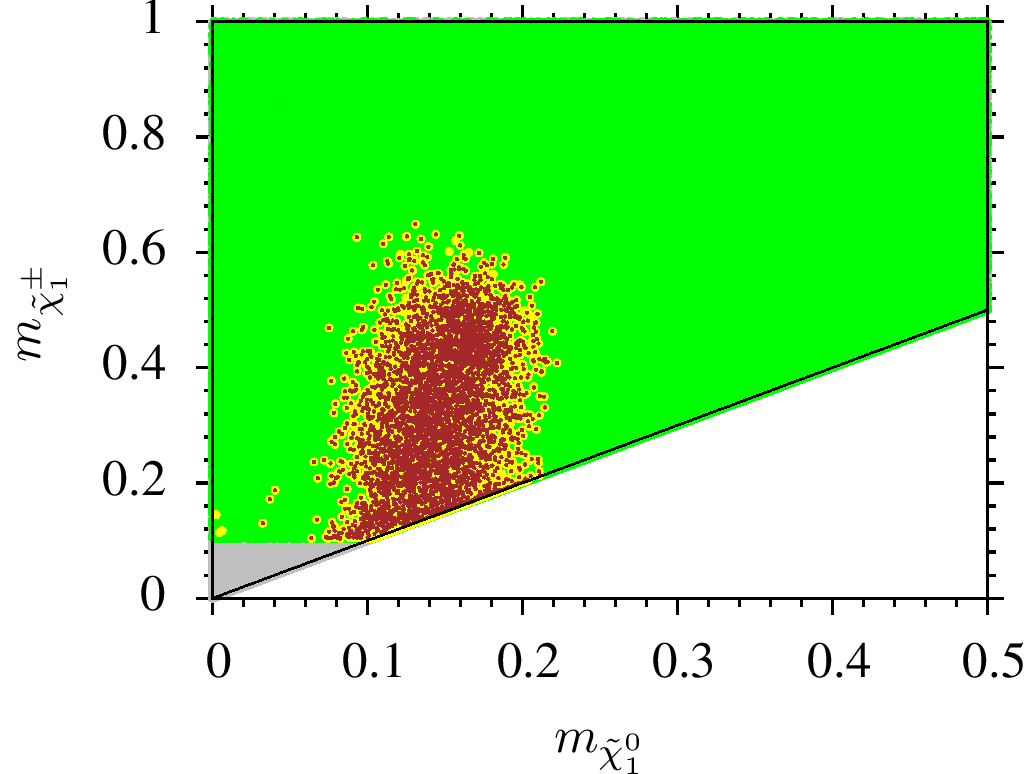}}
\subfigure{\hspace{1.0cm}\includegraphics[scale=1]{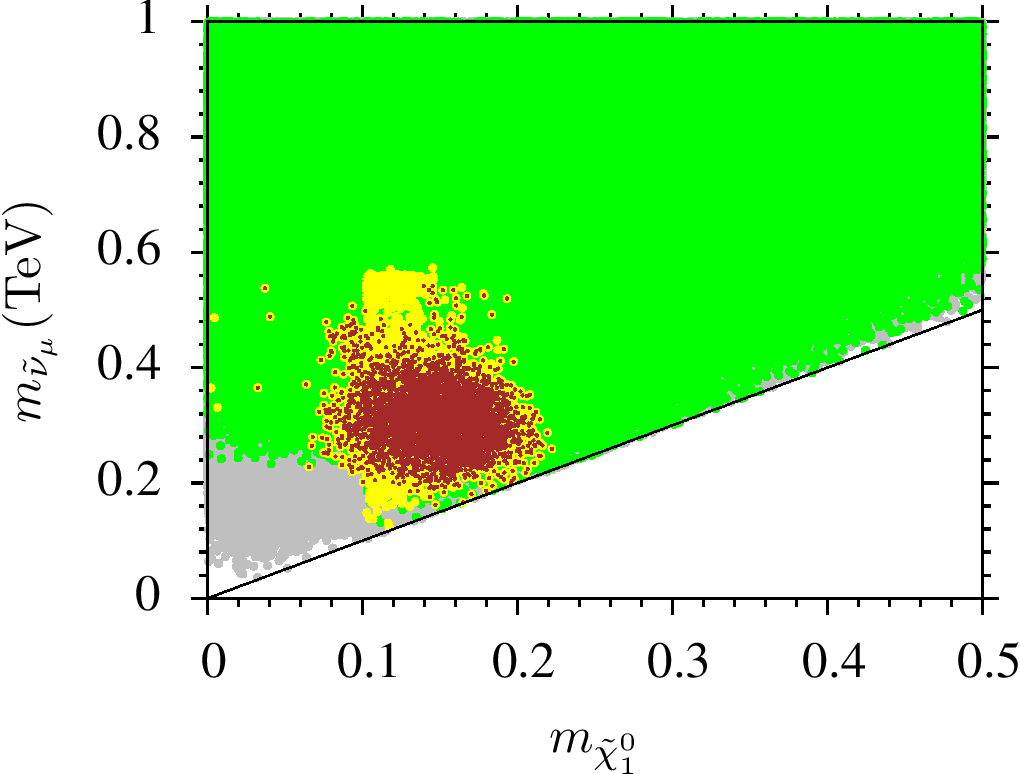}}
\caption{Plots in the $m_{\tilde{\mu}_{R}}-m_{\tilde{\chi}_{1}^{0}}$, $m_{\tilde{\tau}_{1}}-m_{\tilde{\chi}_{1}^{0}}$, $m_{\tilde{\chi}_{1}^{\pm}}-m_{\tilde{\chi}_{1}^{0}}$ and $m_{\tilde{\nu}_{\mu}}-m_{\tilde{\chi}_{1}^{0}}$ planes. Color coding is the same as in Figure \ref{fig:3}.}
\label{fig7}
\end{figure}

Figure \ref{fig6} shows the gaugino mass ratios, and the gaugino and sfermion masses at $M_{{\rm GUT}}$ in the $\Delta a_{\mu}-M_{3}/M_{1}$, $\Delta a_{\mu}-M_{3}/M_{2}$, $\Delta a_{\mu}-M_{2}/M_{1}$, $\Delta a_{\mu}-M_{2}$, $\Delta a_{\mu}-M_{3}$ and $\Delta a_{\mu}-m_{16}$ planes, with the color coding  as in Figure \ref{fig:1}.  {We find} $|M_{3}|/|M_{1}| \gtrsim 1$, while $|M_{3}|/|M_{2}| \gtrsim 3.4$. In contrast to the previous case, the 1$\sigma$ limit on $g-2$ requires the ratio $|M_{2}|/|M_{1}| \lesssim 2.5$. As seen from the $\Delta a_{\mu}-M_{2}$ panel, muon $g-2$ prefers $M_{2}\lesssim 1$ TeV, while it allows only large values of $M_{3}$ ($\gtrsim 2$ TeV) dictated by the 125 GeV Higgs boson requirement. As expected, the sfermion masses turn out to be light, and the $\Delta a_{\mu}-m_{16}$ plane shows that the $m_{16}$ can be as heavy as $\sim 700$ GeV.

\begin{figure}[]
\begin{center}
\includegraphics[scale=1]{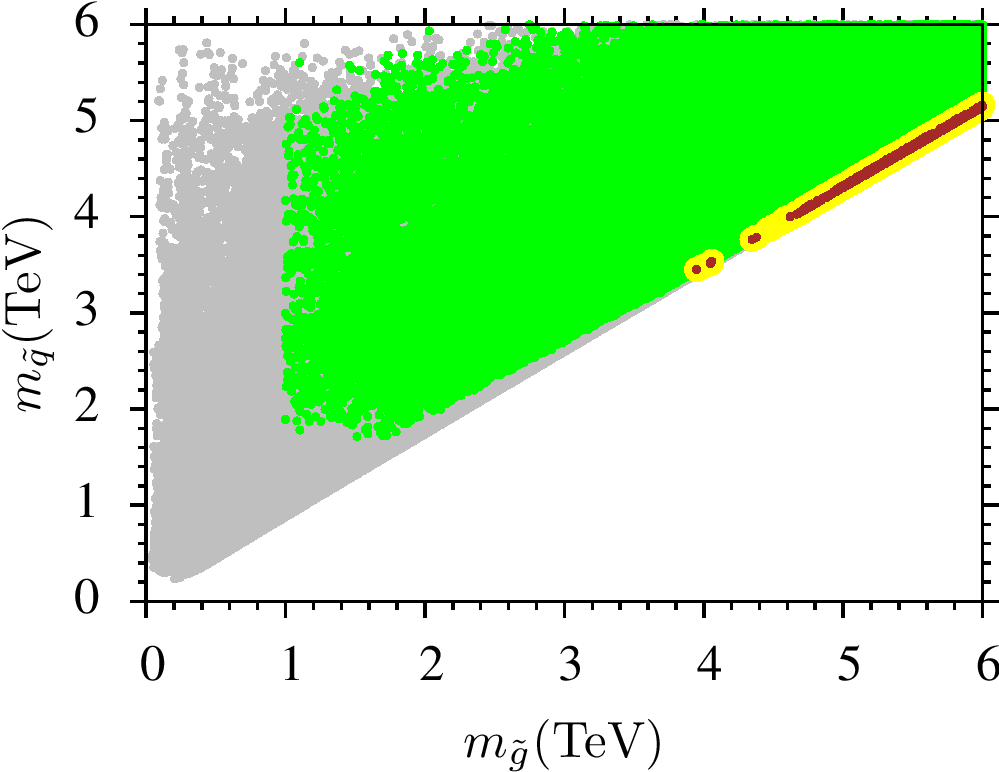}
\end{center}
\caption{Plot in the $m_{\tilde{q}}$-$m_{\tilde{g}}$ plane. Color coding is the same as in Figure \ref{fig:3}.}
\label{fig8}
\end{figure}

Figure \ref{fig7} displays the possible coannihilation channels in the $m_{\tilde{\mu}_{R}}-m_{\tilde{\chi}_{1}^{0}}$, $m_{\tilde{\tau}_{1}}-m_{\tilde{\chi}_{1}^{0}}$, $m_{\tilde{\chi}_{1}^{\pm}}-m_{\tilde{\chi}_{1}^{0}}$ and $m_{\tilde{\nu}_{\mu}}-m_{\tilde{\chi}_{1}^{0}}$ panels. { Since this} scenario allows only light LSP solutions, the coannihilation channels require {the appropriate NLSP to be sufficiently  light and nearly} degenerate with the LSP. The coannihilation scenarios { are similar to those in the} previous section. On the other hand, there is no solution {corresponding to the} A-resonance, while sneutrino-neutralino coannihilation channel is possible in this scenario.
Figure \ref{fig8} shows the result for the squarks and gluino spectra, and we find a heavy spectrum for the colored sparticles ($m_{\tilde{q}} \gtrsim 3$ TeV and $m_{\tilde{g}} \gtrsim$ 4 TeV), similar to the scenario in  {the previous section.}

Table \ref{tab:2} lists three benchmark points for this scenario that satisfy all the constraints described in Section \ref{constraintsSection}. The colored sparticles are all quite heavy while the  {sleptons are light ($\sim$ few hundred GeV)}. Points 1-3 respectively correspond to smuon-neutralino, stau-neutralino and chargino-neutralino coannihilation channels.

\begin{table}[]\hspace{-1.0cm}
\centering
\begin{tabular}{|c|ccc|}
\hline
\hline
                 & Point 1 & Point 2 & Point 3  \\

\hline
$m_{16}$        & 309.2 & 456.2 & 382.3  \\
$M_{1}$          & 497.7 & 427.6 & 425.7   \\
$M_{2} $       & 720.5 & 442.1 & 276.4  \\
$M_{3} $       & 4610 & 4724 & 3030  \\
$\tan\beta$      & 10.5 & 16.1 & 15.4  \\
$A_0/m_{16}$      & -0.16 & -0.03 & 2.37  \\
$m_{10}$          & 1280 & 241.5 & 391.1  \\
$m_t$            & 173.3  & 173.3 & 173.3  \\
\hline
& & &  \\
$\Delta a_{\mu}$  & $ \mathbf{ 20.7\times 10^{-10} } $ & $ \mathbf{22.6\times 10^{-10} } $ & $ \mathbf{ 21.1\times 10^{-10} } $  \\ & & &  \\

\hline
$m_h$            & \textbf{123.1} & \textbf{123.8} & \textbf{122.1}  \\
$m_H$            & 4868 & 4823 & 3165   \\
$m_A$            & 4837 & 4792 & 3145  \\
$m_{H^{\pm}}$    & 4869 & 4824 & 3166  \\

\hline
$m_{\tilde{\chi}^0_{1,2}}$
                 & \textbf{187.4}, 537.4 & \textbf{154.5}, 295.2 & \textbf{161.6}, \textbf{175.9} \\

$m_{\tilde{\chi}^0_{3,4}}$
                 & 4600, 4600 & 4819, 4819 & 3135, 3135  \\

$m_{\tilde{\chi}^{\pm}_{1,2}}$
                 & 541.8, 4558 & 297.4, 4775 & \textbf{177.5}, 3107  \\

$m_{\tilde{g}}$  & 9153 & 9399 & 6206  \\
\hline $m_{ \tilde{u}_{L,R}}$
                 & 7784, 7806 & 7997, 8021  & 5320, 5337 \\
$m_{\tilde{t}_{1,2}}$
                 & 6693, 7316 & 6927, 7518 & 4630, 5019 \\
\hline $m_{\tilde{d}_{L,R}}$
                 & 7784, 7806  & 7997, 8027 & 5321, 5341 \\
$m_{\tilde{b}_{1,2}}$
                 & 7279, 7764 & 7480, 7955 & 4987, 5296 \\
\hline
$m_{\tilde{\nu}_{1,2}}$
                 & 330.5 & 291.7 & 302.7 \\
$m_{\tilde{\nu}_{3}}$
                 &  354.2 & 342.6 & 313.6 \\
\hline
$m_{ \tilde{e}_{L,R}}$
                & 510.3, \textbf{196} & 487, 389.5 & 392.3, 372.4 \\
$m_{\tilde{\tau}_{1,2}}$
                & 221.4, 470.1  & \textbf{180.3}, 544.3 & 212.1, 456.8  \\
\hline

$\sigma_{SI}({\rm pb})$
                & $0.95\times 10^{-13}$ & $ 0.26\times 10^{-15} $ & $ 0.21\times 10^{-12} $\\

$\sigma_{SD}({\rm pb})$
                & $0.11\times 10^{-9}$ &$ 0.88\times 10^{-10} $ & $ 0.59\times 10^{-9}$ \\

$\Omega_{CDM}h^{2}$
                &  0.11 & 0.12 & 0.09 \\

\hline
\hline
\end{tabular}
\caption{Masses  in this table are in GeV units.
All  points  yield $\Delta a_{\mu}$ in Eq. (\ref{bound2}) within $1\sigma$, and satisfy
 the sparticle mass and B-physics constraints described in Section 3..
 Points 1-3 respectively correspond to smuon-neutralino, stau-neutralino, chargino-neutralino coannihilation channels.
\label{tab:2}
}

\end{table}

\section{ Conclusion \label{conclusions}}

We have explored two classes of supersymmetric models with nonuniversal gaugino masses at $M_ {\rm GUT}$ in order to resolve the muon $g-2$ anomaly encountered in the Standard Model. In both models we find that the resolution of this anomaly is compatible with the presence of a SM-like Higgs boson of mass 125-126 GeV, and the relic LSP neutralino density is compatible with the WMAP dark matter bounds. The Higgs mass bound requires that the colored sparticles are quite heavy, $\gtrsim 3$ TeV, but the sleptons including the smuons can be an order of magnitude or so lighter ($\gtrsim 200$ GeV.)


\section*{Acknowledgments}

We would like to thank Adeel Ajaib for very  useful discussions.
This work is supported in part by the DOE Grant No. DE-FG02-12ER41808.  This work used the Extreme Science
and Engineering Discovery Environment (XSEDE), which is supported by the National Science
Foundation grant number OCI-1053575. I.G. acknowledges support from the  Rustaveli
National Science Foundation  No. 03/79


\end{document}